\documentclass[english,aps,prx,superscriptaddress,twocolumn,address,showacknowledgments,longbibliography]{revtex4-1}
\usepackage[latin9]{inputenc}
\usepackage{color}
\usepackage{amsmath}
\usepackage{amssymb}
\usepackage{graphicx}
\usepackage{esint}
\usepackage{babel}
\begin{document}

\title{Pseudomagnetic fields for sound at the nanoscale}

\author{Christian Brendel}

\affiliation{Institute for Theoretical Physics, University of Erlangen-Nürnberg,
Staudtstr. 7, 91058 Erlangen, Germany}

\affiliation{Max Planck Institute for the Science of Light, Günther-Scharowsky-Stra{ß}e
1/Bau 24, 91058 Erlangen, Germany}

\author{Vittorio Peano}

\affiliation{Institute for Theoretical Physics, University of Erlangen-Nürnberg,
Staudtstr. 7, 91058 Erlangen, Germany}

\author{Oskar Painter}

\affiliation{Institute for Quantum Information and Matter and Thomas J. Watson,
Sr., Laboratory of Applied Physics, California Institute of Technology,
Pasadena, USA }

\author{Florian Marquardt}

\affiliation{Institute for Theoretical Physics, University of Erlangen-Nürnberg,
Staudtstr. 7, 91058 Erlangen, Germany}

\affiliation{Max Planck Institute for the Science of Light, Günther-Scharowsky-Stra{ß}e
1/Bau 24, 91058 Erlangen, Germany}
\begin{abstract}
There is a growing effort in creating chiral transport of sound waves.
However, most approaches so far are confined to the macroscopic scale.
Here, we propose a new approach suitable to the nanoscale which is
based on pseudo-magnetic fields. These fields are the analogue for
sound of the pseudo-magnetic field for electrons in strained graphene.
In our proposal, they are created by simple geometrical modifications
of an existing and experimentally proven phononic crystal design,
the snowflake crystal. This platform is robust, scalable, and well-suited
for a variety of excitation and readout mechanisms, among them optomechanical
approaches.
\end{abstract}
\maketitle
A novel trend has emerged recently in the design of mechanical systems,
towards incorporating topological ideas. These ideas promise to pave
the way towards transport along edge-channels that are either purely
uni-directional \cite{Nash2015} or helical \cite{Susstrunk2015}
(i.e. with two ``spins'' moving in opposite directions), as well
as the design of novel zero-frequency boundary modes \cite{Kane2013,Paulose2015}.
The first few experimental realizations \cite{Susstrunk2015,Paulose2015,Nash2015,Xiao2015,He2015arXiv,RocklinarXiv2015,Chen2014}
and a number of theoretical proposals \cite{Kane2013,Yang2015,Khanikaev2015,Kariyado2015,Bertoldi2015,Pal2016,Fleury2016,Mei2016,Chen2016,Rocklin2016,sussman2016}
involve macroscopic setups. These include coupled spring systems \cite{Prodan2009,Bertoldi2015,Susstrunk2015,Nash2015,Kariyado2015,Pal2016}
and circulating fluids \cite{Yang2015,Khanikaev2015,Chen2016,Lu2016}
for a review \cite{Susstrunk2016}. These designs represent important
proof-of-principle demonstrations of topological acoustics and could
open the door to useful applications at the macroscopic scale. However,
they are not easily transferred to the nanoscale, which would be even
more important for potential applications. 

The first proposal for engineered chiral sound wave transport at the
nanoscale has been put forward in Ref.~\cite{Peano2015}: an appropriately
patterned slab illuminated by a laser with a suitably engineered wavefront
realizes a Chern insulator for sound. The laser drive in \cite{Peano2015}
breaks the time-reversal symmetry, enabling uni-directional topologically
protected transport. On the other hand, there would be clear practical
advantages of a design that operates without any drive and, at the
same time, in a simple nanoscale geometry. By necessity, this must
result in helical transport, with two counter-propagating species
of excitations. 

There are two important classes in this regard: (i) topological insulators,
and (ii) pseudo-magnetic fields. A first idea for (i) at the nanoscale
was put forward recently in Ref. \cite{Mousavi2015}. By contrast,
in the present paper, we show how to engineer arbitrary spatial pseudo-magnetic
field distributions for sound waves in a purely geometry-based design.
In addition (and again in contrast to \cite{Mousavi2015}), it turns
out that our design can be implemented in a platform that has already
been realized and reliably operated at the nanoscale, the snowflake
phononic crystal \cite{SafaviNaeini2014}. That platform has the added
benefit of being a well-studied optomechanical system, which, as we
will show, can also provide powerful means of excitation and readout.
The mechanical pseudo-magnetic fields are analogous to the pseudo-magnetic
fields for electrons propagating on the curved surface of carbon nanotubes
\cite{Kane1997} and in strained graphene \cite{Manes2007,Guinea2010,Levy2010,Low2010}.
Pseudo-magnetic fields mimic real magnetic fields\textcolor{black}{,}
but have opposite sign in the two valleys of the graphene band structure
and, thus, do not break time-reversal symmetry. In the past, this
concept has already been successfully transferred to a photonic waveguide
system \cite{Rechtsman2013}. 

Besides presenting our nanoscale design, we also put forward a general
approach to pseudo-magnetic fields for Dirac quasiparticles based
on the smooth breaking of the ${\cal C}_{6}$ point group and translational
symmetries. Our scheme is especially well suited to patterned engineered
materials such as phononic and photonic crystals. It ties into the
general efforts of steering sound in acoustic metamaterials at all
scales \cite{Maldovan2013,Popa2011,Khelif_Book,Ma2015}. 

\begin{figure*}
\includegraphics[width=2\columnwidth]{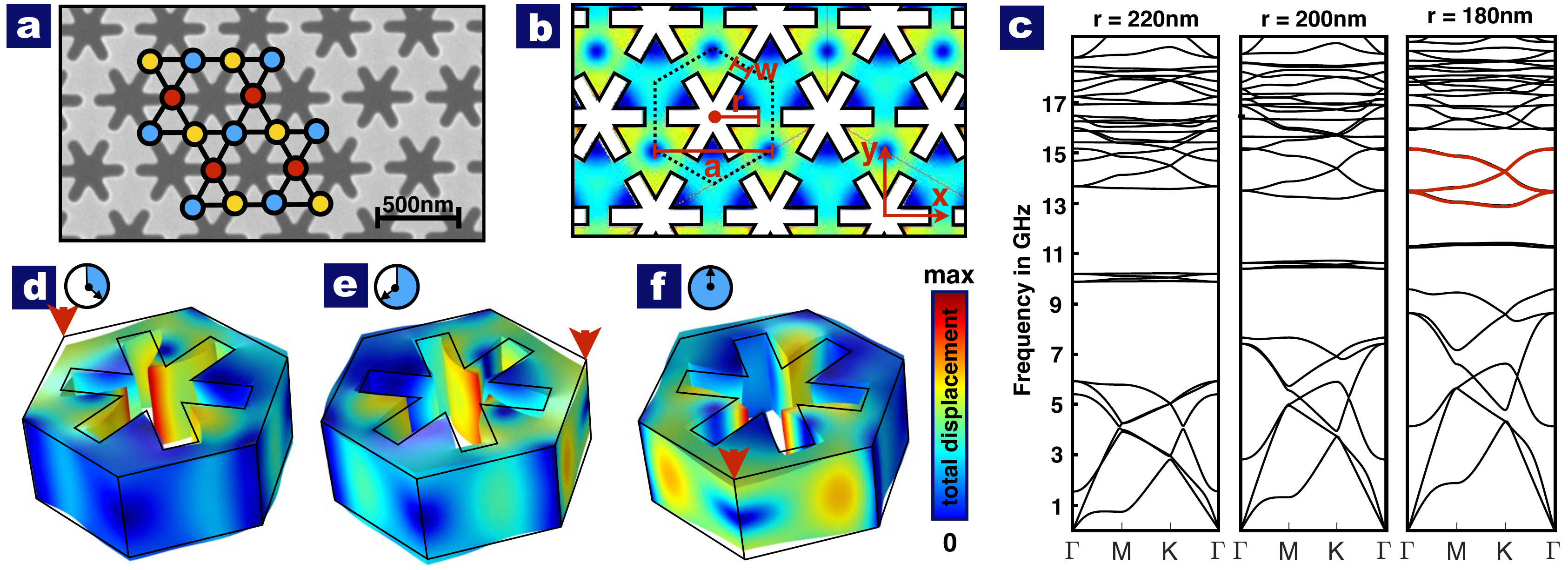}\caption{\label{fig:1}Snowflake crystal: geometry, band structure and displacement
fields. (a) Image of a silicon snowflake phononic crystal. It is formed
by upwards and downwards triangles connected by links. The links form
a Kagome lattice whose Bravais sublattices are marked by circles of
different colors. (b) Top view of the field $|\boldsymbol{\psi}_{m_{s},\tau}(\mathbf{x},z)|^{2}$
(time-averaged square displacement) for one normal mode at the Dirac
point ($m_{s}=-1$ and valley $\tau=-1$). The dashed line indicates
the border of the Wigner-Seitz cell. The relevant geometric parameters
are indicated. (c) \textcolor{black}{Phononic band structure ($z$-symmetric
modes) }for three values of the snowflake radius $r$\textcolor{black}{,
in silicon.} The band structure is computed using a finite-element
solver (COMSOL). The slab thickness is 220nm and the remaining geometry
parameters, c.f. panel (b), are (a,w)=(500,75)nm. For $r=180$nm,
a group of three bands (plotted in red) is separated by complete band
gaps from the remaining bands. The triplet is well fitted by a Kagome
lattice model and exhibits Dirac cones at the high symmetry points
$\mathbf{K}$ and $\mathbf{K'}$ (only $\mathbf{K}$ is shown). At
the cone tip, the degenerate normal modes have quasi-angular momentum
$m_{s}=\pm1$. (d\textendash f) Snapshots of the mechanical displacement
field (absolute value encoded in the color scale) for the mode with
quasi-angular momentum $m_{s}=-1$ and valley $\tau=-1$, within a
Wigner-Seitz cell. Subsequent snapshots are taken after one third
of a period (cf. the clocks). Each snapshot is the anti-clockwise
rotation by a $2\pi/3$ angle around the snowflake center of the previous
snapshot (cf. the arrows). }
\end{figure*}
\emph{Dirac equation and gauge fields \textendash{}} The $2D$ Dirac
Hamiltonian in the presence of a gauge field $\mathbf{A}(\mathbf{x})$
reads (we set the Planck constant and the charge equal to one) \cite{Neto2009}

\begin{equation}
\hat{H}_{D}=m\hat{\sigma}_{z}+v(p_{x}-A_{x}(\mathbf{x}))\hat{\sigma}_{x}+v(p_{y}-A_{y}(\mathbf{x}))\hat{\sigma}_{y}.\label{eq:DiracHamiltonian}
\end{equation}
Here, $m$ is the mass, $v$ is the Dirac velocity, and $\hat{\sigma}_{x,y,z}$
are the Pauli matrices. For zero mass and a constant gauge field ($m=0$
and $\mathbf{A(\mathbf{x})}=\mathbf{A}$), the band structure forms
a Dirac (double) cone, where the top and bottom cones touch at the
momentum $\mathbf{p}=\mathbf{A}$. 

In a condensed matter setting, the Dirac Hamiltonian describes the
dynamics of a particle in a honeycomb lattice, or certain other periodic
potentials, within a quasimomentum valley, i.e. within the vicinity
of a lattice high-symmetry point in the Brillouin zone. In this context,
$\mathbf{p}$ is the quasi momentum counted off from the relevant
high-symmetry point. Here, we are interested in a scenario where the
Dirac Hamiltonian Eq.~(\ref{eq:DiracHamiltonian}) is defined in
two different valleys mapped into each other via the time-reversal
symmetry operator ${\cal T}$. This scenario is realized in graphene,
where the Dirac equation is defined in the two valleys around the
symmetry points $\mathbf{K}$ and $\mathbf{K'}$. For charged particles,
the gauge field $\mathbf{A}(\mathbf{x})$ usually describes a real
magnetic field $\mathbf{B}$, where $\mathbf{B}=\nabla\times\mathbf{A}$.
In this case, the time-reversal symmetry ${\cal T}$ is broken. For
$m=0$ and a constant magnetic field,  the Dirac cones break up in
a series of flat Landau levels \cite{Neto2009},
\begin{equation}
E_{n}={\rm sign}(n)\sqrt{|n|}\omega_{c},\quad\omega_{c}=v\sqrt{2B},\label{eq:Llevels}
\end{equation}
where $n\in Z$ and $\omega_{c}$ is the cyclotron frequency. The
presence of a physical edge then leads to topologically protected
gapless edge states in each valley. For a real magnetic field, the
edge states in the two valleys have the same chirality. However, here,
we will be interested in the case of engineered pseudo-magnetic fields,
where the gauge field $\mathbf{A}(\mathbf{x})$ does not break the
${\cal T}$-symmetry. In this case, $\mathbf{B}=\nabla\times\mathbf{A}$
must have opposite sign in the two different valleys to preserve the
${\cal T}$-symmetry. It is clear that as long as one can focus on
a single valley, the nature of the magnetic field (real or pseudo
magnetic field) does not play any role. This holds true also in the
presence of boundaries. For a given valley and gauge potential $\mathbf{A}$,
exactly the same edge excitations will emerge in the presence of a
pseudo- or a real magnetic field. The nature of the magnetic field
only becomes apparent when the eigenstates belonging to inequivalent
valleys are compared. When time-reversal is preserved (pseudo magnetic
field), each edge state in one valley has a time-reversed partner
with opposite velocity in the other valley. Thus, the edge states
induced by a pseudo magnetic field are not chiral but rather helical.

\emph{Dirac phonons in the snowflake phononic crystal} \textendash{}
FEM mechanical simulations of a silicon thin-film snowflake crystal
are presented in Figure \ref{fig:1}. \textcolor{black}{Throughout
this work, we restrict our attention to the modes that are even under
the mirror symmetry $(x,y,z)\rightarrow(x,y,-z)$, i.e. the $z$-symmetric
modes.} The mechanical band structure is shown in Fig. \ref{fig:1}c.
It features a large number of Dirac cones at the high-symmetry point
$\mathbf{K}$. Each cone  has a time-reversed partner at the point
$\mathbf{K}'$ (not shown). These pairs of Dirac cones are robust
structures: when the radius of the snowflake is varied, they are shifted
in energy (and can possibly cross other bands) but the top and bottom
cones always touch at the corresponding high-symmetry point, see panel
c. In other words, the mass $m$ and the gauge field $\mathbf{A}$
are always zero in the corresponding Dirac Hamiltonian. In order to
generate the desired gauge field, it is necessary to modify the pattern
of holes in a way that breaks the symmetries of the crystal (see discussion
below). 

In preparation of this, we use the snowflake radius as a knob to engineer
a pair of Dirac cones which are spectrally well isolated from other
bands and have a large velocity. The snowflake crystal can be viewed
as being formed by an array of triangular membranes arranged on a
honeycomb lattice and connected through links (see Figure \ref{fig:1}b).
In principle, we could choose a situation where the links are narrow
(large snowflake radius $r$), such that all the groups of bands are
spectrally well isolated. However, then the Dirac velocities tend
to be small. For wider links (smaller $r$), the motion of the adjacent
edges of neighboring triangular membranes becomes strongly coupled.
This give rise to normal modes where such adjacent sides oscillate
in phase, resulting in large displacements of the links. We note that
these links are arranged on a Kagome lattice, see Figure \ref{fig:1}a.
This observation explains the emergence (see r=180nm plot of Fig.
1e) of a group of three bands, separated from the remaining bands
by complete band gaps, and supporting large velocity Dirac cones.
 The triplet of isolated bands can be well fitted by a Kagome lattice
tight-binding model with nearest-neighbor and next-nearest neighbor
hopping. The Kagome lattice model would be entirely sufficient to
guide us in the engineering of the desired gauge fields. However,
we prefer to pursue a more fundamental and general approach based
on the symmetries of the underlying snowflake crystal.

\emph{Identifying the Dirac pseudo-spin by the symmetries} \textendash{}
The snowflake thin-film slab crystal has ${\cal D}_{6h}$ point group
symmetry. If we restrict our attention to the $z$-symmetric modes,
the remaining point group is ${\cal C}_{6v}$ (six-fold rotations
about the snowflake center and mirror symmetries about the vertical
planes containing a lattice basis vector). The degeneracies underpinning
the Kagome Dirac cones as well as the other robust cones in Fig. \ref{fig:1}b
are usually referred to as essential degeneracies. They are preserved
if the point group includes at least the ${\cal C}_{6}$ symmetries
(six-fold rotational symmetry about the snowflake center) or the ${\cal C}_{3v}$
symmetries (three-fold rotations and mirror symmetries about three
vertical planes containing a lattice unit vector). The point group
here contains both groups but, for concreteness, our explanation will
focus on the ${\cal C}_{6}$ symmetry. It is useful to think of the
${\cal C}_{6}$ symmetry as a combination of a ${\cal C}_{3}$ (three-fold)
symmetry group and a ${\cal C}_{2}$ (two-fold) symmetry group. 
The three-fold rotations ${\cal C}_{3}$ belong to the group of the
high-symmetry points $\mathbf{K}$ and $\mathbf{K}'$ (they leave
each of these point invariant modulo a reciprocal lattice vector).
As a consequence, at these points, the eigenmodes can be chosen to
be eigenvectors of the ${\cal C}_{3}$ rotations with quasi-angular
momentum $m_{{\rm s}}=0,\pm1$. The essential degeneracies come about
 because the eigenvectors with non-zero quasi-angular momentum $m_{s}$
come in quadruplets (a degenerate pair in each inequivalent valley),
mapped into each other via the time-reversal symmetry operator ${\cal T}$
and the rotation $\hat{R}(\pi)_{z}$ by $180^{o}$ about the snowflake
center (the sole non-trivial element of the ${\cal C}_{2}$ group).
If we denote the members of a quadruplet by $\boldsymbol{\psi}_{m_{s},\tau}(\mathbf{x},z)$,
where $m_{s}=\pm1$ and $\tau=\pm1$ indicates the valley and $z$
is the vertical coordinate, we have
\[
\mathbf{\boldsymbol{\psi}}_{m_{s},\tau}={\cal T}\mathbf{\boldsymbol{\psi}}_{-m_{s},-\tau}=\hat{R}(\pi)_{z}\mathbf{\boldsymbol{\psi}}_{m_{s},-\tau}={\cal T}\hat{R}(\pi)_{z}\mathbf{\boldsymbol{\psi}}_{-m_{s},\tau}.
\]
Note that both ${\cal T}$ and $R(\pi)$ change the sign of the quasimomentum
and, thus, of $\tau$. However, only ${\cal T}$ changes the sign
of the quasi-angular momentum.

The Dirac Hamiltonian (\ref{eq:DiracHamiltonian}) for a given valley
$\tau$ is obtained by projecting the underlying elasticity equations
onto a two-dimensional Hilbert space spanned by the normal modes 
\begin{equation}
\boldsymbol{\psi}_{\mathbf{p},m_{s},\tau}(\mathbf{x},z)=e^{i\mathbf{p}\cdot\mathbf{x}}\boldsymbol{\psi}_{m_{s},\tau}(\mathbf{x},z),
\end{equation}
and by identifying $-m_{s}=\pm1$ with the eigenvalues of the $\hat{\sigma}_{z}$
matrix (see Appendix \ref{AppB}). In other words, the quasi-angular
momentum $m_{s}$ plays the role of the Dirac pseudospin. A mass term
is forbidden because states with equal quasimomentum and opposite
quasi-angular momentum are mapped into each other by the symmetry
${\cal T}\hat{R}(\pi)_{z}$, $\boldsymbol{\psi}_{\mathbf{p},m_{s},\tau}(\mathbf{x},z)={\cal T}\hat{R}(\pi)_{z}\boldsymbol{\psi}_{\mathbf{p},-m_{s},\tau}(\mathbf{x},z)$.
A gauge field $\mathbf{A}$ is also forbidden because it would couple
states with different quasi-angular momentum at the symmetry point.

In our phononic Dirac system, the eigenstates $\boldsymbol{\psi}_{m_{s},\tau}(\mathbf{x},z)$
are three-dimensional complex vector fields. They yield the displacement
fields, 
\begin{equation}
\mathbf{u}_{m_{s},\tau}(\mathbf{x},z,\phi)={\rm Re}[\exp(-i\phi)\mathbf{\boldsymbol{\psi}}_{m_{s},\tau}(\mathbf{x},z)]
\end{equation}
where $\phi$ is the phase of the oscillation. In this classical setting,
$|\boldsymbol{\psi}_{m_{s},\tau}(\mathbf{x},z)|^{2}$ can be interpreted
as the square displacement averaged over one period, $|\boldsymbol{\psi}_{m_{s},\tau}(\mathbf{x},z)|^{2}=\pi^{-1}\int_{0}^{2\pi}d\phi|\mathbf{u}_{m_{s},\tau}(\mathbf{x},z,\phi)|^{2}$.
We note that the field $|\boldsymbol{\psi}_{m_{s},\tau}(\mathbf{x},z)|^{2}$
is invariant under threefold rotations about three inequivalent rotocenters:
the center of the snowflake and the centers of the downwards and upwards
triangles (see Figure \ref{fig:1}b). Three snapshots of the instantaneous
displacement field for the state with $m_{s}=-1$ and $\tau=-1$ are
shown in Fig. \ref{fig:1}d--f. By definition of a quasi-angular momentum
eigenstate with $m_{s}=-1$, when the phase $\phi$ varies by $2\pi/3$
(after one third of a period), the instantaneous displacement field
is simply rotated clockwise by the same angle. When the valley is
known, the quasi-angular momentum (which here play the role of the
pseudospin) can be directly read off from a single snapshot based
on the position of the nodal lines. For $m_{s}\tau=1$ $(m_{s}\tau=-1)$,
they are located at the center of the downwards (upwards) triangles,
cf. Fig. \ref{fig:1}b,d--f.  (For a detailed explanation see Appendix
\ref{AppB}). Below, we will take advantage of our insight on the
symmetries of the pseudospin eigenstates to engineer a local force
field which selectively excites uni-directional waves. 

\begin{figure}
\includegraphics[width=1\columnwidth]{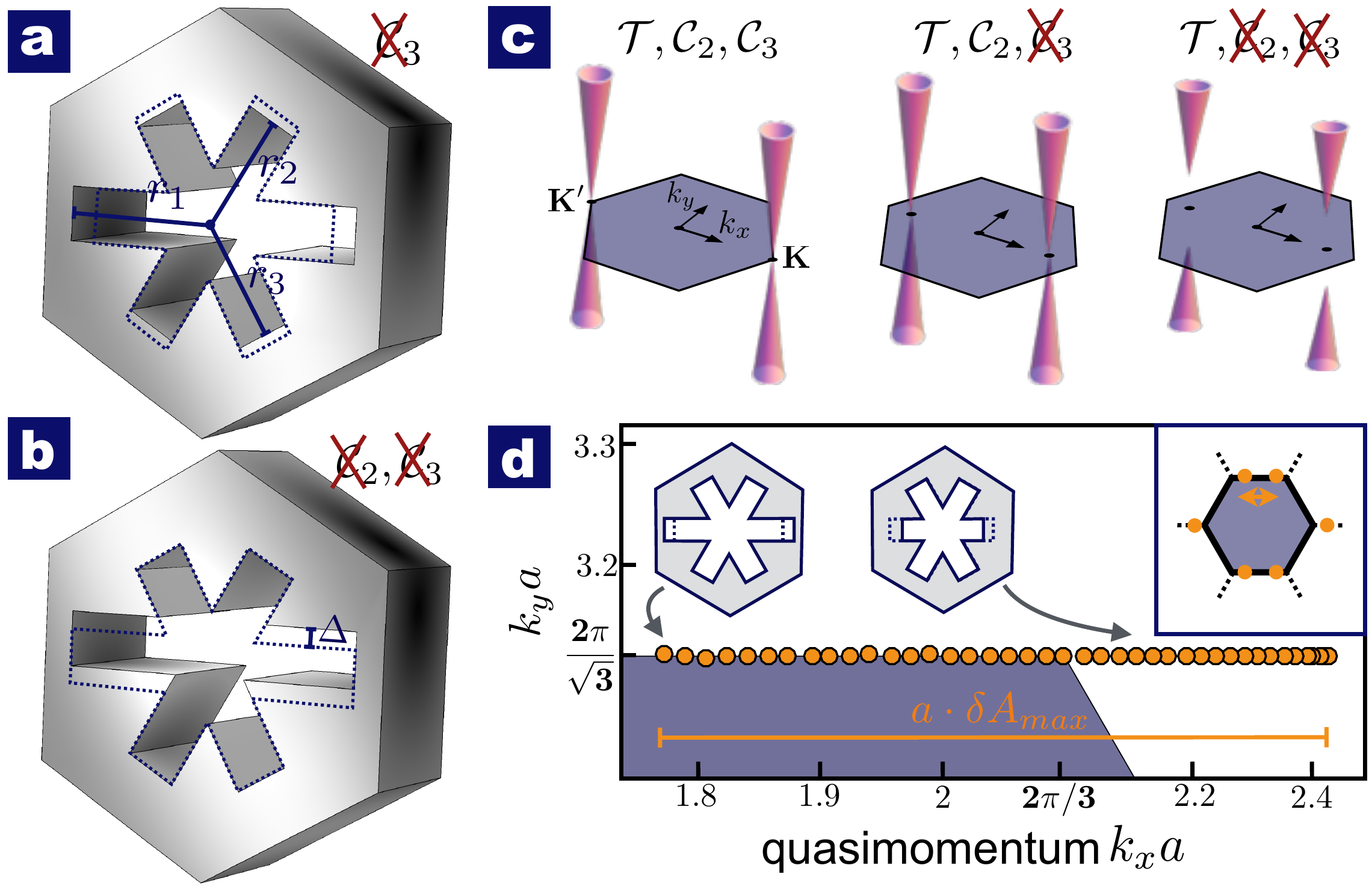}\caption{\label{fig:2}Snowflake geometry and Dirac cones: (a) Geometry of
the snowflake unit cell, depicting a situation with a broken three-fold
rotational symmetry ${\cal C}_{3}$ but preserved two-fold rotational
symmetry ${\cal C}_{2}$ ($r_{1}\protect\neq r_{2}=r_{3}$). (b) Geometry
where both ${\cal C}_{3}$ and ${\cal C}_{2}$ symmetries are broken
(a snowflake arm is displaced vertically by $\Delta$). (c) Resulting
shape of the Dirac cones. Breaking ${\cal C}_{3}$, while time-reversal
symmetry $\mathcal{T}$ and ${\cal C}_{2}$ are maintained, leads
to gapless cones displaced from the high-symmetry points. When also
the ${\cal C}_{2}$ symmetry is broken, a band gap (mass) separating
the upper and lower cones appears. (d) Displacement of the Dirac cone
for the valley $\tau=-1$, for the geometry in panel (a) when $r_{1}$
is varied from $160$nm to $200$nm, while $r_{2}=r_{3}$ and $\bar{r}=(r_{1}+r_{2}+r_{3})/3=180\text{nm}.$}
\end{figure}

\emph{Pseudo-magnetic fields and symmetry-breaking \textendash{}}
A crucial step towards the engineering of a pseudo-magnetic field
is the implementation of a spatially constant vector potential $\mathbf{A}$
in a translationally invariant system. Afterwards arbitrary magnetic
field distributions can be generated straightforwardly by breaking
the translational invariance smoothly. 

A perturbation that breaks the ${\cal C}_{3}$ symmetry but preserves
the ${\cal C}_{2}$ symmetry will simply shift the Dirac cones, without
opening a gap (see Figure \ref{fig:2}c). This can be identified with
the appearance of a constant gauge field $\mathbf{A}$ in the Dirac
Hamiltonian (\ref{eq:DiracHamiltonian}). As such, the connection
between changes in the microscopic structure of the phononic metamaterial
and the resulting gauge field can be obtained from FEM simulations
by extracting the quasimomentum shift of the Dirac cones. We emphasize
that, in this context, the vector potential $\mathbf{A}$ has the dimension of an inverse
length. 

In the snowflake phononic crystal, we can achieve the desired type
of symmetry breaking (breaking ${\cal C}_{3}$ while preserving ${\cal C}_{2}$)
by designing asymmetric snowflakes formed by arms of different lengths,
$r_{1},$ $r_{2},$ $r_{3}$ (see Figure \ref{fig:2}a). If only one
of the arms is changed, symmetry requires that the vector potential
$\mathbf{A}$ points along that arm as shown in Fig. \ref{fig:2}d.
For the Dirac cones associated with the Kagome lattice, our FEM simulations
show that the cone displacement grows linearly with the length changes,
as long as these remain much smaller than the average arm length $r=(r_{1}+r_{2}+r_{3})/3$.
 In this linear regime, and for a general combination of arm lengths,
$r_{1},$ $r_{2},$ $r_{3}$, we have 
\begin{equation}
\mathbf{A}\approx\tau f(r)\mathbf{d},\quad\mathbf{d}=(r_{1}\mathbf{e}_{1}+r_{2}\mathbf{e}_{2}+r_{3}\mathbf{e}_{3}).\label{eq:vectorpot}
\end{equation}
The unit vectors $\mathbf{e}_{j}$ point into the direction of the
corresponding snowflake arms, $\mathbf{e}_{j}=\cos\theta_{j}\mathbf{e}_{x}+\sin\theta_{j}\mathbf{e}_{y}$,
where $\theta_{j}=2\pi(j-1)/3$. The factor $\tau=\pm1$ appears because
the vector potential has opposite sign in the two valleys as we have
not broken time reversal symmetry. We note that in general changes
of the arm lengths also shift the frequency of the Dirac point. When
the arm lengths are chosen to be position dependent, as is required
for implementing arbitrary magnetic fields, this energy shift will
enter the Dirac equation as a scalar potential $V(\mathbf{x})$, which
may be unwanted. However, our numerical simulations show that we can
keep $V(\mathbf{x})$ approximately constant, by retaining a constant
average arm length $r$. 

\begin{figure}
\includegraphics[width=1\columnwidth]{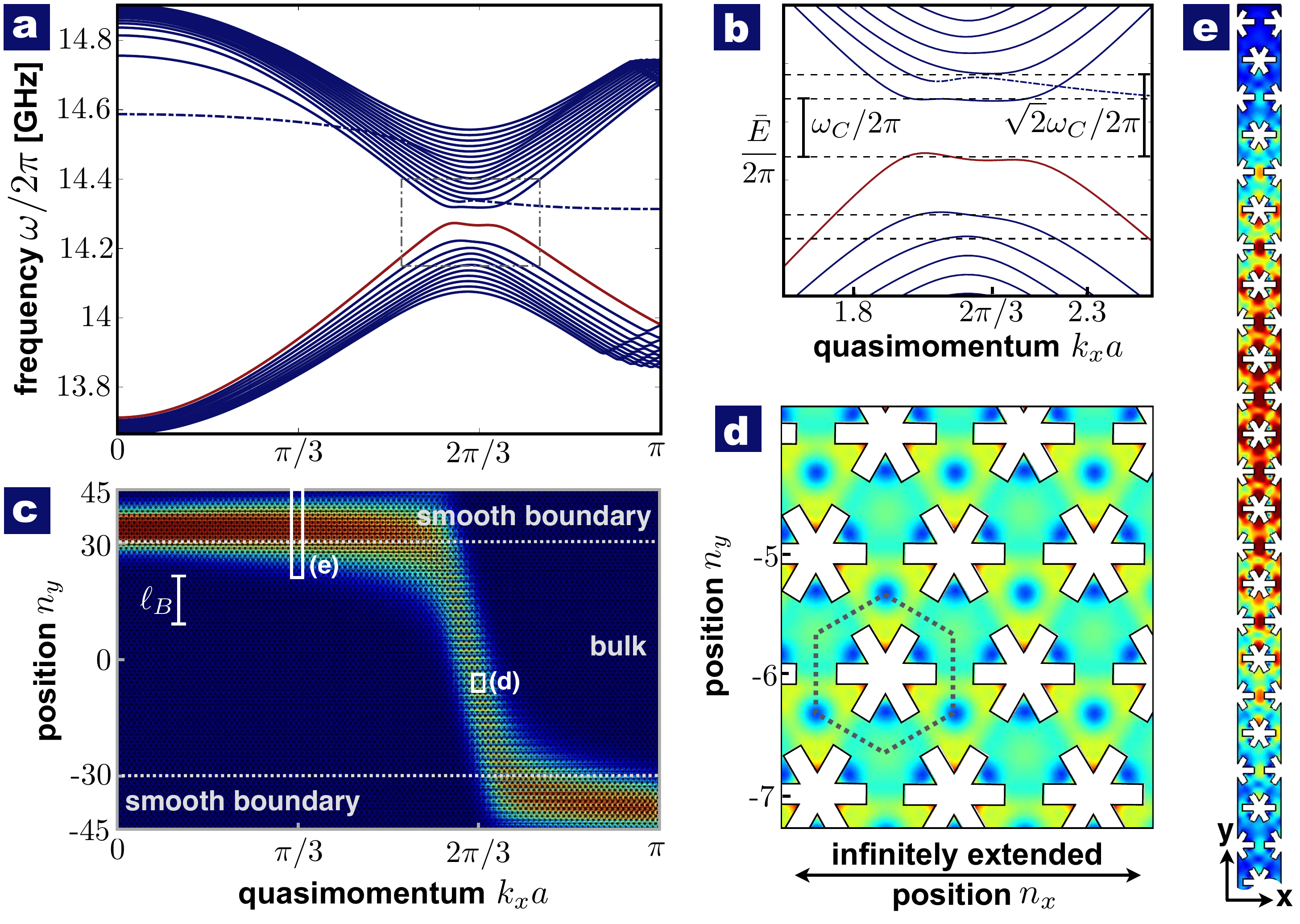}\caption{\label{fig:3}Band structure and displacement fields of a strip in
a constant pseudo-magnetic field: (a) Band structure (only the $z$-symmetric
modes are shown) and (b) zoom-in of the valley $\tau=-1$. The band
structure displays multiple flat Landau plateaus in the vicinity of
the $\mathbf{K}$'-point ($k_{x}a=2\pi/3$). In (b), the dashed lines
indicate the energies of the Landau levels as calculated from the
Dirac equation for mass $m=0$, cf. Eq. (\ref{eq:Llevels}). In (a,b),
the dot-dashed bands correspond to intrinsic non-chiral edge states
located at the physical boundary of the system (see Appendix \ref{AppE}).
(c) Mode shape of the 0-th Landau level and the ensuing edge states
(marked in red in panels a,b) as a function of the quasimomentum $k_{x}$.
The region of the smooth boundaries (where $m\protect\neq0$) is marked
in grey and the magnetic length $\ell_{B}$ in white. (d) Zoom-in
of the displacement field of the $n=0$ Landau level. At the lattice
scale, the displacement field pattern encodes the pseudospin $m_{s}=1$
of the Landau level (cf. Figure \ref{fig:1}f). (e) Zoom-in of the
edge state displacement field.}
\end{figure}

\emph{Phononic Landau levels and Edge States in a Strip \textendash{}
}We can test these concepts by implementing a constant phononic pseudo-magnetic
field in an infinite snowflake crystal strip, where we can directly
test our simplified description against full microscopic simulations.
The strip is of finite width $W$ in the $y$-direction (where $-W/2<y<W/2$).
We can realize the corresponding vector potential in the Landau gauge,
$\mathbf{A}(\mathbf{x})=(-B_{z}y,0,0)$, by varying the length $r_{1}$
along the $x$-axis, while keeping the remaining arm lengths equal,
$r_{2}=r_{3}$. For concreteness we choose $B_{z}>0$ for $\tau=-1$.

The treatment of the boundaries merits special consideration. It turns
out that sharp boundaries are unfavorable, as they give rise to an
extra undesired edge mode that is not related to the quantum Hall
effect physics that we seek to implement. \textcolor{black}{Any smooth
gradient of snowflake parameters near the boundaries of the sample
will lead to well-defined edge states that are spatially separated
from the physical system edge. In general, this could involve both
a potential gradient as well as a gradient in the effective mass (gap).
In our simulations, we will display results obtained for a smooth
mass gradient, whose details do not matter for the qualitative behavior.}
A Dirac mass term appears upon breaking the ${\cal C}_{2}$ symmetry,
which we here choose to do by transversally displacing one of the
snowflake arms, as shown inFig. \ref{fig:2}b, with the displacement
varying smoothly in the interval $W_{{\rm bulk}}/2<|y|<W/2$. 

By changing the snowflake arm lengths we can displace the Dirac cones
only over a finite range of quasimomenta. In our simulations $\delta A_{{\rm max}}\approx0.17\pi/a$,
as shown in Fig. \ref{fig:2}d. Using Eq. (\ref{eq:Llevels}) and
the definition of the magnetic length $\ell_{B}=B^{-1/2}$, we see
that there is a trade-off between the cyclotron frequency $\omega_{c}$
and, thus, the achievable magnetic band gaps and the system size in
the appropriate magnetic units: $\omega_{c}\leq\sqrt{2}v\delta A_{{\rm max}}/w$
where $w=W/\ell_{B}.$ For our FEM simulations we have chosen\textbf{
$w=6.2$}. 

In Fig. \ref{fig:3}, we display the phonon band structure and the
phonon wave functions (mechanical displacement fields) extracted from
finite-element numerical simulations as a function of the quasimomentum
$k_{x}$ along the translationally invariant (infinite) direction.
We display only positive $k_{x}$ because, due to time-reversal symmetry,
both the frequencies and the displacement fields are even functions
of $k_{x}$. In the bulk, we expect to reproduce the well-known physics
of Dirac materials in a constant (pseudo)-magnetic field \cite{Neto2009}.
Indeed, the numerically extracted band structure consists of a series
of flat Landau Levels at energies of precisely the predicted form
$\omega=\bar{E}+E_{n}$ {[}$E_{n}$ is defined in Eq. (\ref{eq:Llevels}){]};
see panel (a) and the zoom-in (b) of Fig. \ref{fig:3}. The Landau
plateaus extend over a quasi momentum interval of width $\delta k_{x}\approx\delta A_{{\rm bulk}}=W_{{\rm bulk}}B$.
Furthermore, in the bulk, we expect the mechanical eigenstates to
be localized states of size $\ell_{B}=B^{-1/2}$ (in the $y$-direction).
Their quasi momentum $k_{x}$ should be related to the position via
$\bar{y}=-[k_{x}-2\pi/(3a)]/B_{z}$. This behavior is clearly visible
in Fig. \ref{fig:3}c, where we show the displacement field of the
central Landau level. A zoom-in of this field (panel d) reveals that,
at the lattice scale, it displays the same intensity pattern as the
bulk pseudospin eigenstate $m_{s}=-1$ shown previously in Fig. \ref{fig:1}f.
This behavior is also predicted by the effective Dirac description
where the central Landau level is indeed a pseudo-spin eigenstate
with $m_{s}=-1$ when the magnetic field $B_{z}$ is positive \cite{Neto2009}.
Note that the pseudo-magnetic field engineered here also gives rise
to a Lorentz force that will curve the trajectory of phonon wavepackets
traveling in the bulk of the sample. The sign of the force is determined
by the valley index $\tau$.

Having demonstrated that we can implement a constant phononic pseudo-magnetic
field, we now argue that our approach with smooth boundaries gives
additional flexibility in the engineering of helical phononic waveguides.
Each Landau level gives rise to an edge state in the region of the
smooth boundaries. The typical behavior of the wavefunction is shown
(for $n=0$) in Figure \ref{fig:3}c. For decreasing quasimomenta
$k_{x}$, an edge state localized on the lower edge smoothly evolves
into a bulk state, and eventually into an edge state localized on
the upper edge. As is clear from Fig. \ref{fig:3}a and \ref{fig:3}b,
these pairs of edge states span the same energy interval. This behavior
clearly leads to the same number of edge states (for a fixed energy)
on both edges. This is crucial in view of engineering smooth helical
transport on a closed loop. We emphasize that the number of states
on the two edges need not necessarily (by symmetry) be equal. Indeed,
graphene in a constant pseudo-magnetic field supports a different
number of edge states on two opposite (sharp) edges \cite{Low2010}.
In our approach, we can tune the number of edge states on each edge
via the mass term. In particular, the behavior of the edge states
originating from the $n=0$ Landau level is sensitive to the sign
of the mass. A negative mass (as in our simulations) drags this Landau
level into the band gap below. Vice versa, a positive mass will drag
it into the band gap above. This behavior is related to the peculiarity
of the Landau level being a pseudospin eigenstate (with $m_{s}=-1$)
and, thus, an eigenstate of the mass term (with eigenvalue $m$),
cf. Eq. (\ref{eq:DiracHamiltonian}).

\begin{figure}
\includegraphics[width=1\columnwidth]{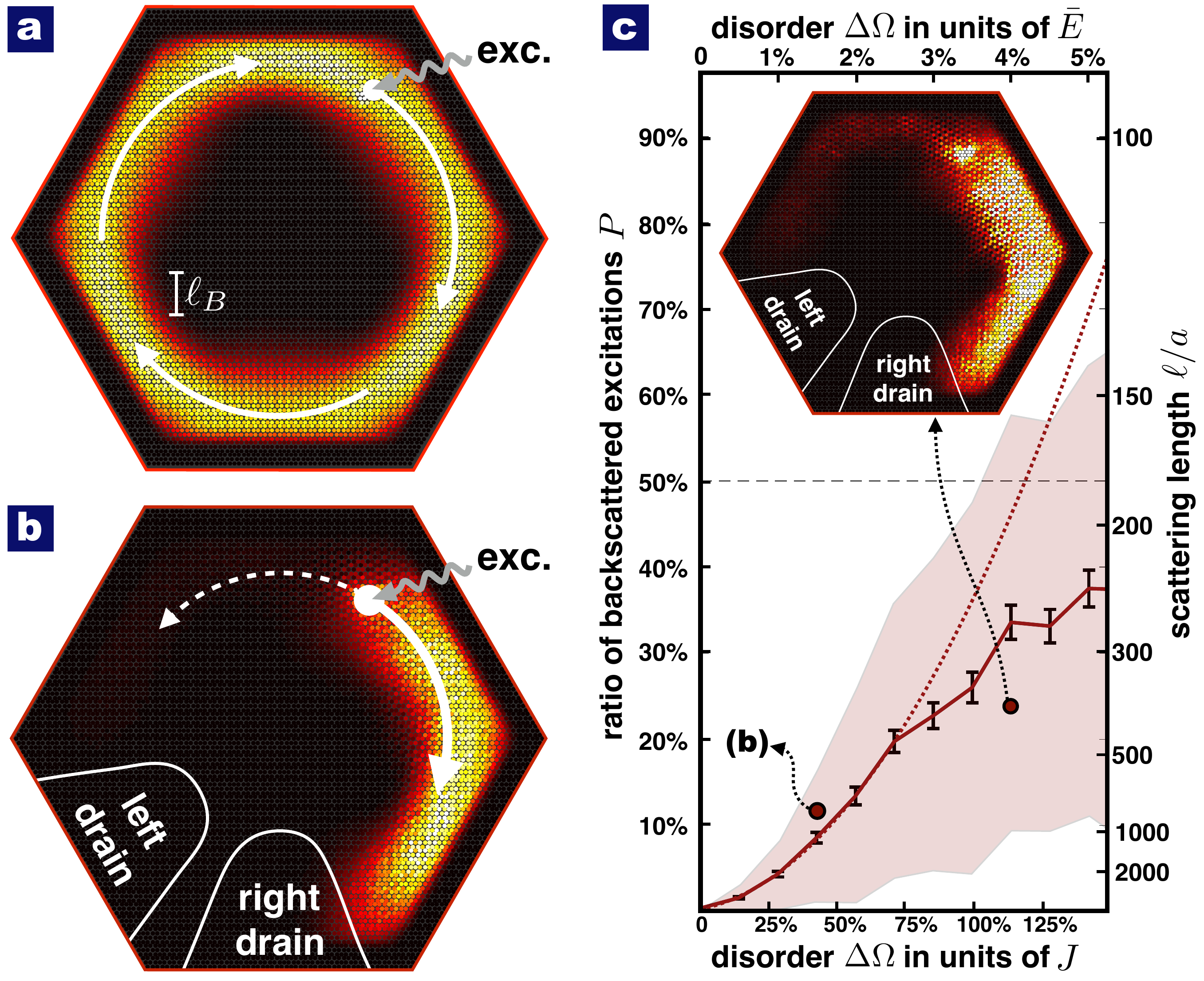}\caption{\label{fig:4}Tight-binding transport simulations in a finite geometry:
The hexagon (height 55 unit cells) comprises an internal bulk area
(height 45 unit cells) and external smooth boundaries. An engineered
oscillating force (at the position indicated by the grey arrow) with
frequency in the middle of a band gap (separating the Landau levels
$n=-1$ and $n=0$) launches clockwise propagating sound waves. (a)
Resulting displacement field (unit-cell resolved). (b) and inset of
(c) Displacement field in the presence of disorder and drains where
the excitation are absorbed. (c) Probability $P$ (averaged over 150
disorder implementations) that an excitation is absorbed in the left
drain as a function of the disorder strength. The estimated error
is represented by error bars. The shaded area represents the values
of $P$ within one standard deviation. In the tight-binding model,
the random onsite energies are equally distributed on an interval
of width $\Delta\Omega$. For weak disorder ($\Delta\Omega/J\lesssim70\%$
where $J$ is the average hopping rate), $P$ is well fitted by a
parabola (dashed line), indicating that the transport is quasi-ballistic.
In this regime, the corresponding scattering length $\ell$ is shown
in the right vertical axis. }
\end{figure}

\emph{Transport in a Finite Geometry and Disorder \textendash{}} Any
pseudo-magnetic field that is realized without explicit time-reversal
symmetry breaking necessarily gives rise to helical transport, where
the chirality depends on some artificial spin degree of freedom, i.e.
the valley. A central question in this regard is the robustness against
short-range disorder. In order to assess this, we have studied numerically
transport in a finite geometry. As presented in Fig. 4, we consider
a sample of hexagonal shape with smooth boundaries in the presence
of a constant pseudo-magnetic field (we choose the symmetric gauge
for the vector potential ${\bf A}$). In this illustrative example,
a local probe drive excites vibrations near the boundaries, as indicated
in Fig. \ref{fig:4}a. Its frequency is chosen to lie inside the bulk
band gap separating the $n=0$ and $n=-1$ Landau levels. In this
band gap, the system supports a pair of counter-propagating helical
edge states belonging to opposite valleys. One can select a propagation
direction by engineering the driving force. In a simple setting, one
could apply a time-dependent force that is engineered to excite only
the pseudo-spin eigenstate $m_{s}=1$ in the valley $\tau=1$. This
can be achieved by exerting forces at the three corners of a Wigner-Seitz
cell, where the eigenstate displays a large vertical displacement,
see Figure \ref{fig:1}b-d. There is a phase delay of $2\pi/3$ between
any two corners while a similar pattern of phase delays but with opposite
signs occur in the other valley. Thus, a force which is modulated
with the right phase delays will selectively drive the valley $\tau=1$
and excite only excitations with a particular chirality. 

It turns out to be most efficient (and entirely sufficient) to implement
the numerical simulations for these rather large finite-size geometries
with the help of a tight-binding model on a Kagome lattice (see Appendix
\ref{appF}). The parameters of that model can be matched to full
FEM simulations that have been performed for the translationally invariant
case.  This allows us to systematically study the effects of disorder.
In the presence of moderate levels of smooth disorder, which does
not couple the two valleys, the nature of the underlying magnetic
field (pseudo vs. real) will not manifest itself and the transport
will largely be immune to backscattering. Here, we focus instead on lattice-scale
disorder which can lead to scattering with large momentum transfer
that couples the two valleys and thereby leads to backscattering.
\textcolor{black}{We emphasize that short-range disorder will, in
practice, introduce backscattering in any purely geometric approach
to acoustic helical transport. In particular, this also includes acoustic
topological insulators, where generic disorder will break the unitary
symmetry that ultimately protects the transport \cite{Lu2014}.} To
quantify the effect of lattice scale disorder, we consider a setup
with two drains, one to the clockwise and one to the counter-clockwise
direction, as shown in Fig. \ref{fig:4}b. In the absence of disorder,
the vibrations travel clockwise (in this example) and are absorbed
in the right drain; only very weak residual backscattering occurs
at the sharp hexagon corners. In the presence of lattice-scale disorder,
a portion of the excitations will be backscattered and subsequently
reach the left drain. In \ref{fig:4}c, we plot the portion $P$ of
excitations absorbed in the left drain, averaged over a large number
of disorder implementations, as a function of the disorder strength.
In the regime of quasi-ballistic transport (for weak enough disorder),
$P$ is proportional to the backscattering rate. Thus, it scales as
the square of the disorder amplitude and can be used to extract the
scattering length $\ell$: $P=d/\ell$, where $d$ is the distance
between source and drain. In current nanoscale snowflake crystal experiments
\cite{SafaviNaeini2014}, the fabrication-induced geometric disorder
is on the order of $1\%$ of the absolute mechanical frequency $\bar{E}$
which corresponds to $25\%$ of the average hopping rate in the tight-binding
model. In that case, our simulations indicate the resulting scattering
lengths $\ell$ to be very large (of the order of more than $1000$
snowflake unit cells).

\emph{Implementation \textendash{}} Since our design is scale invariant,
a variety of different implementations can be easily envisioned. At
the nanoscale, the fabrication of thin-film silicon snowflake crystals
and resonant cavities have already been demonstrated with optical
read-out and actuation \cite{SafaviNaeini2014}. At the macro scale,
the desired geometry could be realized using 3D-laser printing and
similar techniques. A remaining significant challenge relates to the
selective excitation of helical sound waves and the subsequent read
out. In an optomechanical setting, the helical sound waves can be
launched by carefully crafting the applied radiation pressure force.
For the typical dimensions of existing snowflake optomechanical devices
operating in the telecom wavelength band (lattice spacing $a\approx500$nm),
the required force could be engineered using tightly-focused intensity-modulated
laser beams impinging from above on three different snowflake triangles.
The read-out could occur by measuring motionally-induced sidebands
on the the reflection of a laser beam. Although the direct radiation
pressure of the beam will induce rather weak vibrations, they could
still be resolved using optical heterodyning techniques. Alternatively,
in a structure scaled up $10$ times, selected triangles could host
defect mode optical nanocavities. This would boost the radiation pressure
force and the read-out precision by the cavity finesse (see \ref{appG}).
Helical sound waves can then be launched by either directly modulating
the light intensity or a photon-phonon conversion scheme, using a
strong red-detuned drive, with signal photons injected at resonance.
In the micron regime one can benefit from electro-mechanical interactions.
A thin film of conducting material deposited on top of the silicon
slab in combination with a thin conducting needle pointing towards
the desired triangles forms a capacitor. In this setting, an AC voltage
would induce the required driving force. The vibrations could be read
out in the same setup as they are imprinted in the currents through
the needles. In a different electromechanical approach, the phononic
crystal could be made out of a piezoelectric material and excitation
and read out occur via piezoelectric transducers \cite{Balram2016}. 

\emph{Conclusions \textendash{} }We have shown how to engineer pseudo-magnetic
fields for sound at the nanoscale purely by geometrical means in a
well established platform. Our approach is based on the smooth breaking
of the ${\cal C}_{6}$ and the discrete translational symmetry in
a patterned material; it is, thus, of a very general nature and directly
applies to photonic crystals as well. Indeed, the same geometrical
modifications that have led to the pseudomagnetic fields for sound
investigated in our work will also create pseudomagnetic fields for
 light \cite{Rechtsman2013,Lu2014} in the same setup.  Our approach
offers a new paradigm to design helical photonic and phononic waveguides
based on pseudo-magnetic fields. 

\emph{Acknowledgements\textendash{}} V.P., C.B., and F.M. acknowledge
support by the ERC Starting Grant OPTOMECH and by the European Marie-Curie
ITN network cQOM; O.P. acknowledges support by the AFOSR-MURI Quantum
Photonic Matter, the ARO-MURI Quantum Opto-Mechanics with Atoms and
Nanostructured Diamond (grant N00014-15- 1-2761), and the Institute
for Quantum Information and Matter, an NSF Physics Frontiers Center
(grant PHY-1125565) with support of the Gordon and Betty Moore Foundation
(grant GBMF-2644).

\appendix

\section{Kagome-Dirac modes from Another Point of View\label{AppA}}

\begin{figure*}
\includegraphics[width=2\columnwidth]{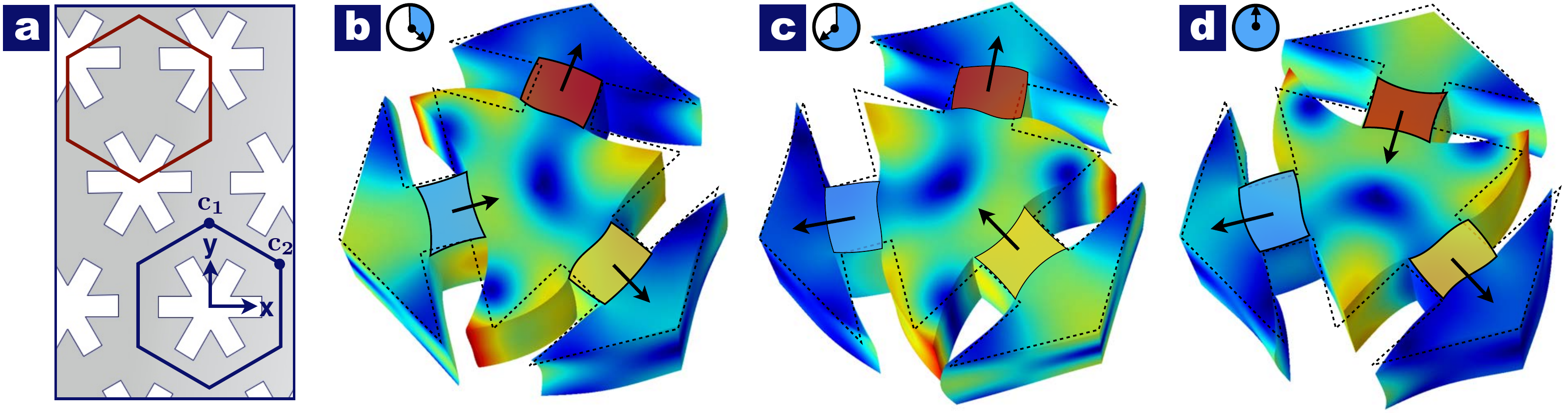}

\caption{\label{fig:Unitcells}Eigenmodes of the silicon snowflake phononic
crystal. (a) Different choices of the unit-cell: While the blue cell
will nicely illustrate the motion of the snowflake itself (cf. Figure
1c), the red one rather emphasizes the motion of the links connecting
the triangles (cf. b-d). (b-d) Displacement field of the same eigenmode
shown in Figure 1c of the main text ($m_{s}=-1$, $\tau=-1$). Subsequent
snapshots are taken after one third of a period (cf. the clocks),
where the arrow indicates the current direction of the links' velocities.
Notice that the quasi-angular momentum about the center of the unit
cell is $m_{1}=1$ (anti-clockwise rotation).}
\end{figure*}

In the main text, we have plotted the normal modes in a Wigner-Seitz
cell around the center of the snowflake. An alternative choice that
highlights better the motion of the links forming the Kagome lattice
is to center the Wigner-size cell around the center of a triangle,
see Figure \ref{fig:Unitcells}a. In Figure \ref{fig:Unitcells}b-d,
we show the same normal mode plotted in Figure 1d-f of the main text
for the different Wigner-Seitz cell. Note that the picture is rotated
anti-clockwise by a $2\pi$/3 angle after one third of a period corresponding
to the quasi-angular momentum about the center of the unit cell $m_{1}=1$
, cf. Eq. (\ref{eq:m1m2}) with $\tau=-1$ and $m_{s}=-1$ .

\section{Explanation of the essential degeneracies and symmetries of the pseudospin
eigenstates\label{AppB}}

Here, we discuss the symmetries of the normal modes $\boldsymbol{\psi}_{m_{s},\tau}$
of the snowflake crystal at the high-symmetry points. Thereby we also
explain the appearance of the essential degeneracies underpinning
the robust Dirac cones, and explain how to identify a pseudospin eigenstate
from the FEM simulation of a strip.

We consider a generic mode $\boldsymbol{\psi}(\mathbf{x},z)$ with
quasimomentum $\mathbf{k}$, 
\[
\hat{T}(\boldsymbol{a})\boldsymbol{\psi}(\mathbf{x},z)=\boldsymbol{\psi}(\mathbf{x}-\boldsymbol{a},z)=e^{-i\mathbf{k}\cdot\boldsymbol{a}}\boldsymbol{\psi}(\mathbf{x},z)
\]
 where $\hat{T}(\boldsymbol{a})$ is a translation by a lattice vector
$\boldsymbol{a}$. As usual we can use the Bloch ansatz 
\begin{equation}
\boldsymbol{\psi}(\mathbf{x},z)=e^{i\mathbf{k}\cdot\mathbf{x}}\boldsymbol{\phi}(\mathbf{x},z),\label{eq:defofu}
\end{equation}
where $\boldsymbol{\phi}(\mathbf{x},z)$ is periodic under discrete
translations, $\boldsymbol{\phi}(\mathbf{x},z)=\boldsymbol{\phi}(\mathbf{x}-\boldsymbol{a},z)$.
We choose the center of the snowflake as the origin of the coordinates,
see Figure \ref{fig:Unitcells}a. By applying a rotation by a angle
$\theta$ about the $z$-axis, we find
\begin{eqnarray}
\hat{R}(\theta)_{z}[\boldsymbol{\psi}(\mathbf{x},z)] & = & R(\theta)_{z}\boldsymbol{\psi}(R(-\theta)\mathbf{x},z)\nonumber \\
 & = & e^{i\mathbf{k}\cdot R(-\theta)\mathbf{x}}R(\theta)_{z}\boldsymbol{\phi}(R(-\theta)\mathbf{x},z)\nonumber \\
 & = & e^{iR(\theta)\mathbf{k}\cdot\mathbf{x}}R(\theta)_{z}\boldsymbol{\phi}(R(-\theta)\mathbf{x},z)\label{eq:rotation}
\end{eqnarray}
where 
\begin{equation}
R(\theta)_{z}=\begin{pmatrix}R(\theta) & 0\\
0 & 1
\end{pmatrix},\quad R(\theta)=\begin{pmatrix}\cos\theta & -\sin\theta\\
\sin\theta & \cos\theta
\end{pmatrix}.\label{eq:2Drot}
\end{equation}
We note that $\hat{R}(\theta)_{z}[\boldsymbol{\psi}(\mathbf{x})]$
has a quasimomentum $\mathbf{k}'=R(\theta)\mathbf{k}$ rotated by
$\theta$. For a triangular lattice, the high-symmetry points $\mathbf{K}$,
$\mathbf{K}'$, and $\boldsymbol{\Gamma}$ have the peculiarity to
be invariant under ${\cal C}_{3}$ rotations. For example, $R(2\pi/3)\mathbf{K}=\mathbf{K}+\mathbf{b}_{1}$
where $\mathbf{b}_{1}$ is a basis vector of the reciprocal lattice.
Thus, applying the rotation $\hat{R}(2\pi/3)_{z}$ to a state with
quasimomentum $\mathbf{k}=\mathbf{K}$, see Eq. (\ref{eq:rotation}),
we find another state with the same quasimomentum, 
\[
\hat{R}(2\pi/3)_{z}[\boldsymbol{\psi}(\mathbf{x},z)]=e^{i\mathbf{K}\cdot\mathbf{x}}\boldsymbol{\phi}'(\mathbf{x},z)
\]
where 
\[
\boldsymbol{\phi}'(\mathbf{x},z)=e^{i\mathbf{b_{1}\cdot\mathbf{x}}}R(2\pi/3)_{z}\boldsymbol{\phi}(R(-2\pi/3)\mathbf{x},z)
\]
is invariant under $2D$ discrete translations, $\boldsymbol{\phi}'(\mathbf{x}+\boldsymbol{a},z)=\boldsymbol{\phi'}(\mathbf{x},z)$.
In other words, $\hat{R}(2\pi/3)_{z}$ (and more in general any ${\cal C}_{3}$
rotation) commutes with the projector $\hat{P}_{\mathbf{K}}$ onto
the states with quasimomentum $\mathbf{K}$. The same holds for $\mathbf{K}'$,
and $\boldsymbol{\Gamma}$. Thus, for each high-symmetry point $\mathbf{k}=\mathbf{K},\mathbf{K}',$
and $\boldsymbol{\Gamma}$, it is possible to find a basis of eigenstates
of the ${\cal C}_{3}$ rotations spanning the sub-Hilbert space of
states with that particular quasimomentum. If the crystal has the
$2D$ discrete translational invariance and the ${\cal C}_{3}$ symmetry,
such a basis can be chosen to be eigenstates of the Hamiltonian. 

In the following, we denote a common eigenstate of the ${\cal C}_{3}$
rotations and the translations by $\boldsymbol{\psi}_{m_{s},\tau}(\mathbf{x})$
where $\tau$ indicates the valley ($\tau=1$ for $\mathbf{k}=\mathbf{K}$
and $\tau=-1$ for $\mathbf{k}=\mathbf{K}'$ ) and $m_{s}$ the quasi-angular
momentum, 
\[
\hat{R}(2\pi/3)_{z}[\boldsymbol{\psi}_{m_{s},\tau}(\mathbf{x},z)]=e^{-im_{s}2\pi/3}\boldsymbol{\psi}_{m_{s,\tau}}(\mathbf{x},z).
\]
From Eq. (\ref{eq:rotation}), we see that in terms of the corresponding
translational invariant field $\boldsymbol{\phi}_{m_{s},\tau}(\mathbf{x},z)$
we have 
\begin{multline}
e^{i\mathbf{\tau b_{1}\cdot\mathbf{x}}}R(2\pi/3)_{z}\boldsymbol{\phi}_{m_{s},\tau}(R(-2\pi/3)\mathbf{x},z)\\
=e^{-im_{s}2\pi/3}\boldsymbol{\phi}_{m_{s},\tau}(\mathbf{x},z)\label{eq:C3def2}
\end{multline}

Next, we show that the eigenstates with non-zero quasi-angular momentum
$m_{s}=\pm1$ at the valleys $\tau=\pm1$ can be organized in quadruplets
which are degenerate if the Hamiltonian has time-reversal symmetry
${\cal T}$ and the full ${\cal C}_{6}$ symmetry. We denote the members
of the quadruplet $\boldsymbol{\psi}_{m_{s},\tau}(\mathbf{x})$. Starting
from an arbitrary state with $m_{s}=1$ or $m_{s}=-1$, the remaining
three members of the quadruplet are (by definition) obtained by applying
the time-reversal symmetry ${\cal T}$ and $\hat{R}(\pi)_{z}$ (a
rotation by $\pi$ about the $z$-axis),
\begin{eqnarray*}
\boldsymbol{\psi}_{-m_{s},-\tau}(\mathbf{x},z) & = & {\cal T}\boldsymbol{\psi}_{m_{s},\tau}(\mathbf{x},z),\\
\boldsymbol{\psi}_{m_{s},-\tau}(\mathbf{x},z) & = & \hat{R}(\pi)_{z}\boldsymbol{\psi}_{m_{s},\tau}(\mathbf{x},z),\\
\boldsymbol{\psi}_{-m_{s},\tau}(\mathbf{x},z) & = & {\cal T}\hat{R}(\pi)_{z}\boldsymbol{\psi}_{m_{s},\tau}(\mathbf{x},z).
\end{eqnarray*}
where ${\cal T}\boldsymbol{\psi}_{m_{s},\tau}(\mathbf{x},z)=\boldsymbol{\psi}_{m_{s},\tau}^{*}(\mathbf{x},z).$
It is straightforward to explicitly check that the states $\boldsymbol{\psi}_{-m_{s},-\tau}(\mathbf{x},z)$,
$\boldsymbol{\psi}_{m_{s},-\tau}(\mathbf{x},z)$, and $\boldsymbol{\psi}_{-m_{s},\tau}(\mathbf{x},z)$
as obtained via the above definitions from $\boldsymbol{\psi}_{m_{s},\tau}(\mathbf{x})$
are indeed eigenstates of the ${\cal C}_{3}$ rotations and the discrete
translations with the appropriate eigenvalues. For the ${\cal C}_{3}$
rotation we have to show that if Eq. (\ref{eq:C3def2}) is assumed
to hold for a specific choice of $m_{s}$ and $\tau$, it will hold
also for the remaining combinations of $m_{s}$ and $\tau$.

Next we discuss the behavior of the states $\boldsymbol{\psi}_{m_{s},\tau}(\mathbf{x})$
under ${\cal C}_{3}$ rotations about the center of the downwards
and upwards triangles, cf. Figure \ref{fig:Unitcells}a,
\[
\mathbf{c}_{1}=\left(0,\frac{a}{\sqrt{3}}\right),\quad\mathbf{c}_{2}=\left(\frac{a}{2},\frac{a}{2\sqrt{3}}\right),
\]
respectively. We note that these points lie at the corners of the
Wigner-Seitz cell around the ${\cal C}_{6}$ rotocenter (in this case
the snowflake center). Thus, as in any ${\cal C}_{6}$ symmetric triangular
lattice, they are threefold rotocenters. The states $\boldsymbol{\psi}_{m_{s},\tau}(\mathbf{x},z)$
are also eigenstates of the ${\cal C}_{3}$ rotations about $\mathbf{c}_{1}$
and $\mathbf{c}_{2}$ (or about any other point belonging to the corresponding
Bravais lattices) with quasi-angular momentum
\begin{eqnarray}
m_{{\rm 1}} & = & (m_{{\rm s}}+\tau+1){\rm mod}3-1,\nonumber \\
m_{{\rm 2}} & = & (m_{{\rm s}}-\tau+1){\rm mod}3-1,\label{eq:m1m2}
\end{eqnarray}
respectively. Here, we use the definition of the function $(n){\rm mod}3$
where $(-1){\rm mod}3=2$. It is easy to verify the above statement
by applying the rotation $\hat{R}(2\pi/3)_{\mathbf{c}_{i},z}$ about
the point $\mathbf{c}_{i}$ to the normal modes $\boldsymbol{\psi}_{m_{s},\tau}(\mathbf{x},z)$,
\begin{multline*}
\hat{R}(2\pi/3)_{\mathbf{c}_{i},z}\boldsymbol{\psi}_{m_{s},\tau}(\mathbf{x},z)\\
=\hat{T}(\mathbf{c}_{i})\hat{R}(2\pi/3)_{,z}\hat{T}(-\mathbf{c}_{i})\boldsymbol{\psi}_{m_{s},\tau}(\mathbf{x},z)\\
=e^{-i(m_{s}+\tau\mathbf{b_{1}}\cdot\mathbf{c}_{i})2\pi/3}\boldsymbol{\psi}_{m_{s},\tau}(\mathbf{x},z).
\end{multline*}
Taking into account that $\mathbf{b_{1}}\cdot\mathbf{c}_{1}=2\pi/3$
and $\mathbf{b_{1}}\cdot\mathbf{c}_{2}=-2\pi/3$ we arrive to Eq.
(\ref{eq:m1m2}). Since $\boldsymbol{\psi}_{m_{s},\tau}(\mathbf{x},z)$
are simultaneous eigenvectors of the ${\cal C}_{3}$ rotations about
all three inequivalent threefold rotocenters of the crystal (the origin,
$\mathbf{c_{1}}$ and $\mathbf{c_{2}}$), the time-averaged squared
displacement field $|\boldsymbol{\psi}_{m_{s},\tau}(\mathbf{x},z)|^{2}$
is invariant under all these symmetry transformations, cf. Figure
1b of the main text. We note that for $m_{s}\tau=1$ ($m_{s}\tau=-1$),
the quasi-angular momentum about the center of the downward (upwards)
triangles $\mathbf{c}_{1}$ ($\mathbf{c}_{2}$ ) is finite, $|m_{1}|=1$
($|m_{2}|=1$), corresponding to a vortex configuration. Thus, the
wavefunction has nodes at these points, cf. Figure 1b of the main
text. 

We note that a generic pseudospin eigenstate, e. g. the zero Landau
level, is the product of a smooth function and the normal mode $\boldsymbol{\psi}_{m_{s},\tau}(\mathbf{x},z)$.
Thus, it will show the same displacement pattern at the lattice scale,
cf. Fig 3d of the main text. Consequently, the pseudo-spin can be
immediately read off from a FEM simulation of a strip (where the valley
is known) just by observing the location of the nodes of $|\boldsymbol{\psi}_{m_{s},\tau}(\mathbf{x},z)|^{2}$
.

\section{Derivation of the Dirac equation in the presence of the $\boldsymbol{C}_{6}$
symmetry\label{AppC}}

Here, we derive the Dirac Hamiltonian for the case where the ${\cal C}_{6}$
symmetry is preserved {[}Equation (1) of the main text with $m=0$
and $\mathbf{A}=0${]}. In each valley ($\tau=\pm1$), we project
the Hamiltonian onto the states $\boldsymbol{\psi}_{\mathbf{p},1,\tau}(\mathbf{x},z)=e^{i\mathbf{p}\cdot\mathbf{x}}\boldsymbol{\psi}_{1,\tau}(\mathbf{x},z)$
and $\boldsymbol{\psi}_{\mathbf{p},-1,\tau}(\mathbf{x},z)=e^{i\mathbf{p}\cdot\mathbf{x}}\boldsymbol{\psi}_{-1,\tau}(\mathbf{x},z)$,
where the quasimomentum $\mathbf{p}$ counted off from the symmetry
point is assumed to be small. For each $\mathbf{p}$ we define the
Pauli matrices according to 
\[
\hat{\sigma}_{z,\mathbf{p}}=|\boldsymbol{\psi}_{\mathbf{p},-1,\tau}\rangle\langle\boldsymbol{\psi}_{\mathbf{p},-1,\tau}|-|\boldsymbol{\psi}_{\mathbf{p},1,\tau}\rangle\langle\boldsymbol{\psi}_{\mathbf{p},1,\tau}|.
\]
From this definition (assuming the usual commutation relations for
the Pauli matrices) we also have
\[
\hat{\sigma}_{+,\mathbf{p}}=(\hat{\sigma}_{x,\mathbf{p}}+i\hat{\sigma}_{y,\mathbf{p}})/2=|\boldsymbol{\psi}_{\mathbf{p},-1,\tau}\rangle\langle\boldsymbol{\psi}_{\mathbf{p},1,\tau}|.
\]
From the band structure calculated by the FEM simulations (without
the pseudomagnetic fields) we see that the eigenenergies are linear
in $|\mathbf{p}|$ close to the relevant symmetry point (they form
a cone). Thus, the Hamiltonian should be, to first approximation,
linear in $\mathbf{p}$. Taking into account that $\boldsymbol{\psi}_{\mathbf{p},-1,\tau}(\mathbf{x},z)={\cal T}\hat{R}(\pi)_{z}\boldsymbol{\psi}_{\mathbf{p},1,\tau}(\mathbf{x},z)$
and that ${\cal T}\hat{R}(\pi)_{z}$ is a symmetry, a mass term (proportional
to $\hat{\sigma}_{z,\mathbf{p}}$ ) is forbidden and the most general
Hamiltonian which is linear in $\mathbf{p}$ has the form 
\begin{equation}
\hat{H}=\bar{E}+\sum_{\mathbf{p}}\mathbf{v}\cdot\mathbf{p}\hat{\sigma}_{+,\mathbf{p}}+h.c.\label{eq:DiracHamalt}
\end{equation}
where $\bar{E}$ is the degenerate energy of the normal modes $\boldsymbol{\psi}_{m_{s},\tau}(\mathbf{x},z)$.
Under the rotation $\hat{R}(2\pi/3)_{z}$ we have
\begin{multline*}
\hat{R}(2\pi/3)_{z}\hat{H}\hat{R}(-2\pi/3)_{z}\\
=\sum_{\mathbf{p}}\mathbf{v}\cdot\mathbf{p}\hat{R}(2\pi/3)_{z}\hat{\sigma}_{+,\mathbf{p}}\hat{R}(-2\pi/3)_{z}+h.c.\\
=\sum_{\mathbf{p}}\mathbf{v}\cdot\mathbf{p}e^{-i2\pi/3}\hat{\sigma}_{+,R(2\pi/3)\mathbf{p}}+h.c.\\
=\sum_{\mathbf{p}}(R(2\pi/3)\mathbf{v})\cdot\mathbf{p}e^{-i2\pi/3}\hat{\sigma}_{+,\mathbf{p}}+h.c.
\end{multline*}
Since the Hamiltonian is invariant under ${\cal C}_{3}$ rotations
we must have
\[
R(2\pi/3)\mathbf{v}=e^{i2\pi/3}\mathbf{v}.
\]
From Eq. (\ref{eq:2Drot}), we find $\mathbf{v}=v(1,-i)$ where $v$
is the slope of the cones. By plugging into Eq. (\ref{eq:DiracHamalt})
and projecting onto a single quasimomentum we obtain the Dirac equation
(1) of the main text {[}for $m=0$ and $\mathbf{A}=0${]}.

\section{Details of the numerical calculations of the pseudomagnetic fields\label{sec:FEM}}

In this section we provide additional details about the numerical
calculations performed with the COMSOL finite-element solver, thereby
guiding through the computation of the movements of the Dirac cones
in reciprocal space and the construction of the resulting pseudo magnetic
field for phonons in a strip configuration. In all calculations the
material is assumed to be silicon (Si) with Young's modulus of $170\,\text{GPa}$,
mass density $2329\,\text{kg/m}^{3}$ and Poisson's ratio $0.28$.

\begin{figure}
\includegraphics[width=1\columnwidth]{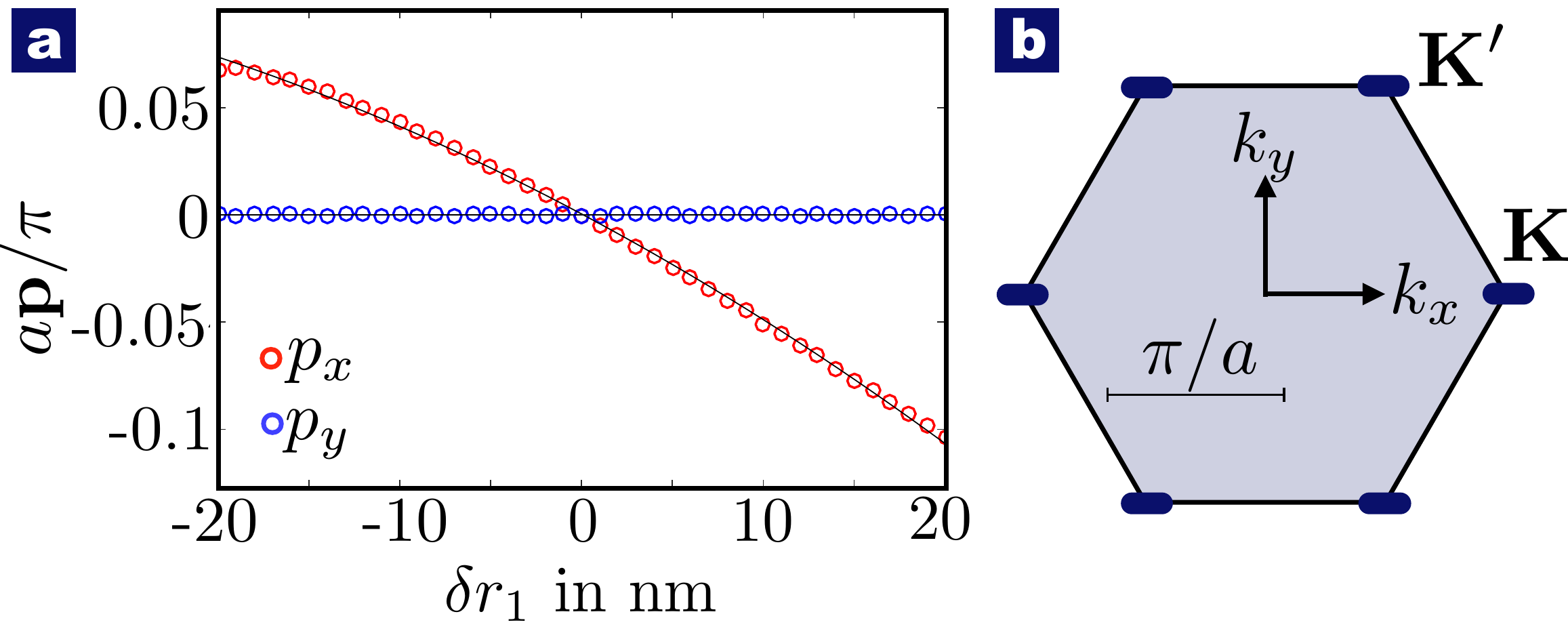}

\caption{\label{fig:QuanitaticeDiracDisplacement}Displacement of the Dirac
cones in reciprocal space due to a broken $\mathcal{C}_{3}$-symmetry
(cf. Figure 2a of the main text). Here the snowflakes arm lengths
$r_{i}=\bar{r}+\delta r_{i}$ are changed with $\delta r_{1}\protect\neq\delta r_{2}=\delta r_{3}$,
while $\delta r_{1}+\delta r_{2}+\delta r_{3}=0$ in order to keep
the average radius constant at $\bar{r}=180\,\text{nm}$. (a) depicts
the shift of the Dirac cone ${\bf p}$ at ${\bf k={\bf K'}}$ ($\tau=\text{-}1$
valley) computed by the COMSOL finite element solver for different
arm length $\delta r_{1}$ (red and blue circles). For the other valley
$(\tau=1$) the cones move in the opposite direction, which builds
the foundation of a pseudo magnetic field for phonons that does not
break the time-reversal symmetry. The data can be very well fitted
by a parabola (solid black lines). (b) Overview of the Brillouin zone,
with the blue bars indicating the range of the Dirac cones position
that can be achieved by a variation of the snowflake arm length by
$\delta r_{1}=[-20;+20]\text{nm}$. }
\end{figure}

Breaking the 3-fold rotational symmetry ($\mathcal{C}_{3}$-symmetry)
of the snowflake geometry but maintaining its inversion symmetry ($\mathcal{C}_{2}$-symmetry),
displaces the Dirac cones from the high symmetry points ${\bf K}$
and ${\bf K}'$, but does not gap the system (cf. Figure 2c of the
main text). This effect is depicted in Figure \ref{fig:QuanitaticeDiracDisplacement},
which shows the motion of the Dirac cones, as an effect of the broken
$\mathcal{C}_{3}$-symmetry. Thereby the radius of the horizontally
orientated snowflake arm is varied by $\delta r_{1}$ ($r_{1}=\bar{r}+\delta r_{1}$),
which displaces the Dirac cones in the $k_{x}$-direction. Generally
this would also shift the Dirac cone's energy. In order to avoid that,
the remaining two snowflake arms are varied equivalently by $\delta r_{2}=\delta r_{3}$
in such a way that the average radius is kept fixed at $\bar{r}=(r_{1}+r_{2}+r_{3})/3=180\,\text{nm}$
(i.e $\delta r_{1}+\delta r_{2}+\delta r_{3}=0$).

\begin{figure}
\includegraphics[width=1\columnwidth]{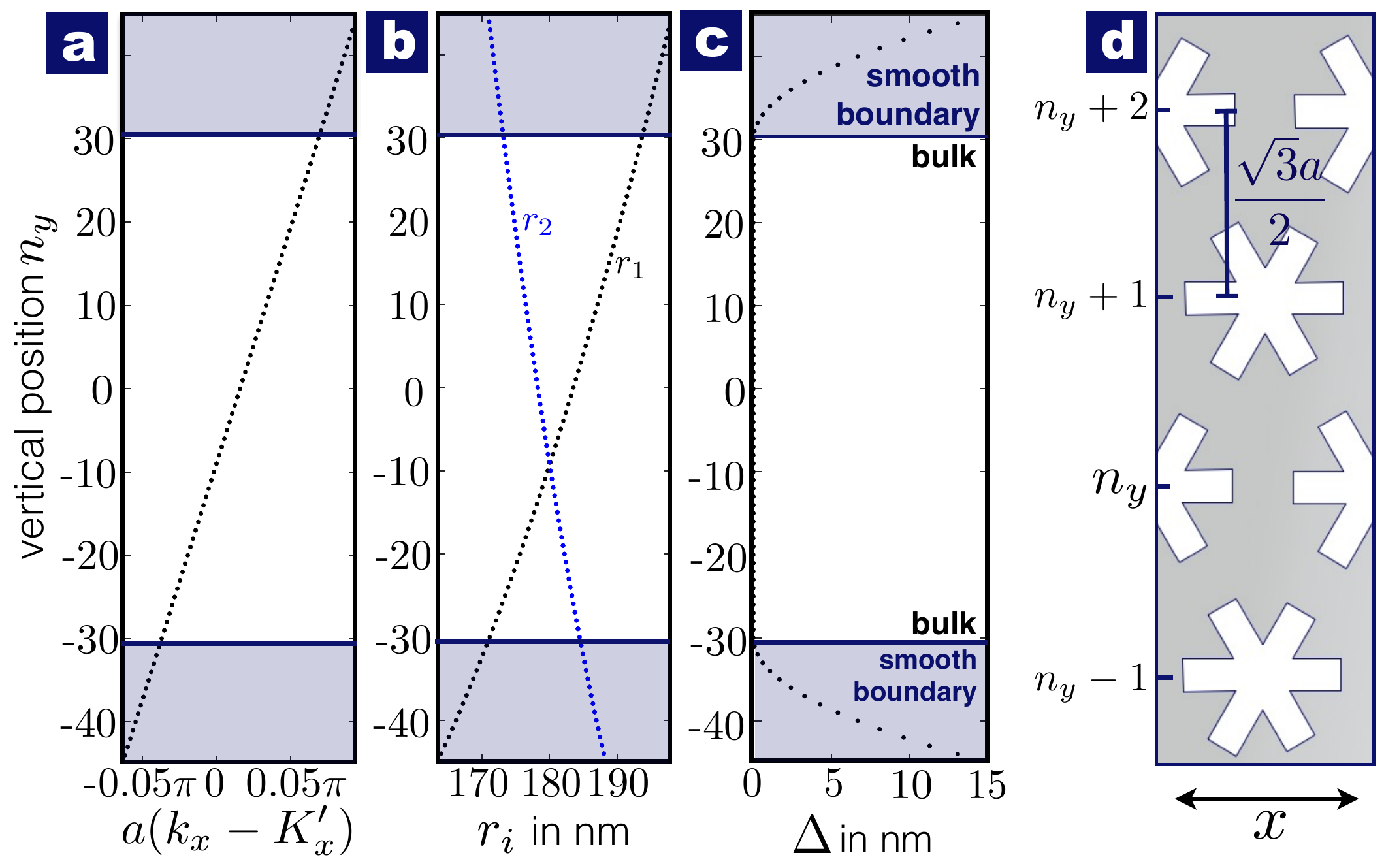}

\caption{\label{fig:StripeDetails}Details about the strip configuration (cf.
Figure 3 of the main text ). (a) depicts the displacement of the Dirac
cone in the $\tau=-1$ valley ($k_{x}=K_{x}'=2\pi/3a$) as a function
of the vertical position $n_{y}$ in the strip, which forms the basis
of a pseudo-magnetic field for phonons. Those shifts arise from the
broken $\mathcal{C}_{3}$-symmetry, due to the variation of the snowflakes'
radii $r_{1}$ and $r_{2}=r_{3}$ (cf. Figure 2a of the main text).
The exact values of $r_{1}$ and $r_{2}$ in dependence of the position
$n_{y}$ can be seen in (b). Displacing one of the snowflake arms
by $\Delta$ (cf. Figure 2b) locally breaks the $\mathcal{C}_{2}$-symmetry,
which forms smooth boundaries. In (c) the displacement $\Delta$ is
depicted in dependence of $n_{y}$. (d) A small fraction of the stipe's
unit cell. The whole strip comprises 61 snowflakes in the bulk area
$(|n_{y}|\le30)$ and additional 30 snowflakes to form the smooth
boundaries ($30<|n_{y}|\le$45). As it is infinitely extended in the
x-direction a Floquet periodic boundary condition is implemented at
the left and right boundaries, whereas the upper and lower boundaries
are kept fixed (${\bf u}=0$).}
\end{figure}

To engineer a constant pseudo magnetic field in a strip configuration,
that is infinitely extended in the x-direction, the snowflakes need
to be designed properly: The B-field is given by $B_{z}=\partial A_{y}/\partial x-\partial A_{x}/\partial y$,
whereas the vector potential for a given valley is directly related
to the shift of the Dirac cones ${\bf A}={\bf p}$. As the strip is
periodically extended in the x-direction the vector field is not allowed
to vary along this direction (i.e.$\partial A_{y}/\partial x=0$),
while $A_{x}$ must depend linearly on the vertical position $y$
(i.e. $\partial A_{x}/\partial y=const.$), in order to have a constant
magnetic field (cf. Figure \ref{fig:StripeDetails}a). Using the relation
between the shift of the Dirac cones and the variation of the radii
(quadratic fit in Figure \ref{fig:QuanitaticeDiracDisplacement}a),
the radii of the snowflakes can be calculated in dependence of their
position in the strip (cf. Figure \ref{fig:StripeDetails}b). In addition
to that, we want to engineer smooth boundaries by opening a mass gap,
which is done by breaking two-fold rotational symmetry at the edges
of the sample. This can be achieved by displacing one of the snowflake
arms by $\Delta$ (cf. Figure 2a of the main text), with $\Delta=0$
in the bulk region while it smoothly increases in the vicinity of
the sample's edge (cf. Figure \ref{fig:StripeDetails}c). 

\section{Edge States at the Physical Boundary\label{AppE}}

\begin{figure*}
\includegraphics[width=2\columnwidth]{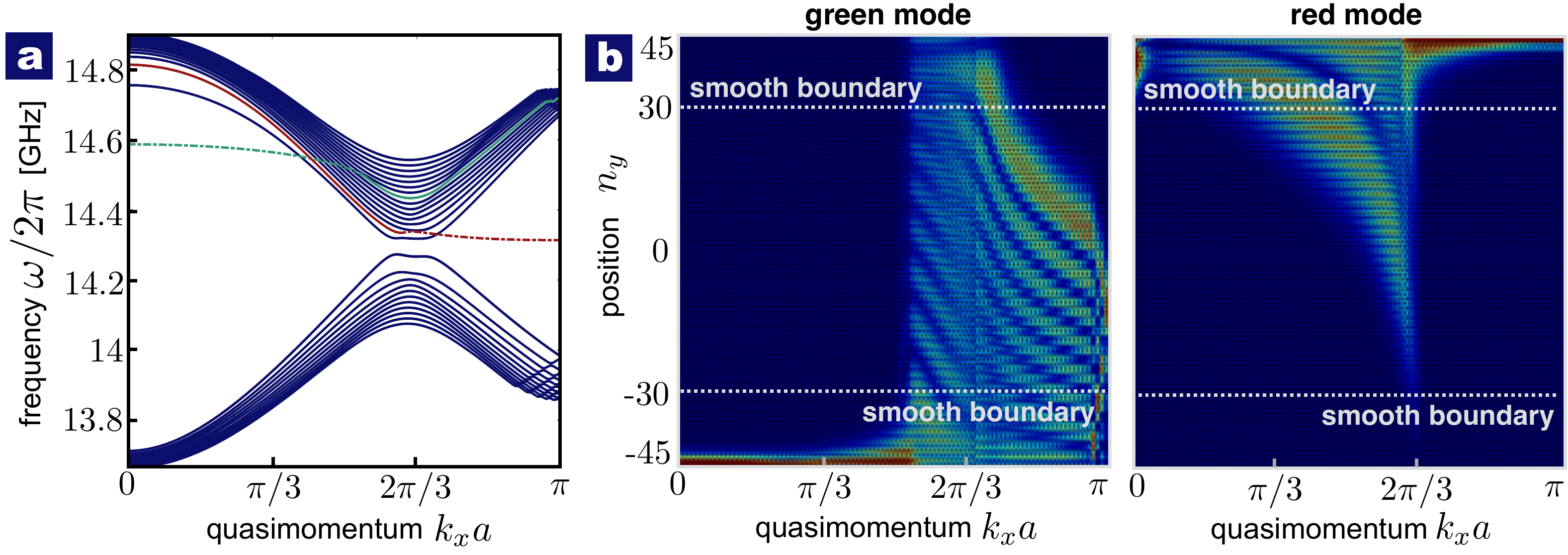}

\caption{\label{fig:edegStates}Intrinsic edge modes at the physical boundaries.
(a) Band structure of the strip configuration discussed in the main
text and in Section \ref{sec:FEM}. The bands highlighted in green
and red feature modes that are highly confined at the physical boundary
of the sample (dashed parts). (b) Displacement field of the red and
green mode. Due to the small slope of their bands for corresponding
quasimomenta, the velocities become zero. Therefore, the edge states
are stationary and do not show any transport properties. }
\end{figure*}

In the Figure \ref{fig:edegStates}, we investigate the intrinsic
edge states that appear at the physical edge of the strip and that
will be present even in the absence of pseudo-magnetic fields. The
relevant bands are highlighted as colored lines in the band structure
of the strip, see panel a. The corresponding displacement fields are
shown in panel b. Note that the edge states are defined only on a
finite portion of the Brillouin zone (where the bands are plotted
as dashed lines) and smoothly transform into bulk modes in the remaining
quasimomentum range.

\section{Tight Binding Model on the Kagome lattice\label{appF}}

\begin{figure}
\includegraphics[width=1\columnwidth]{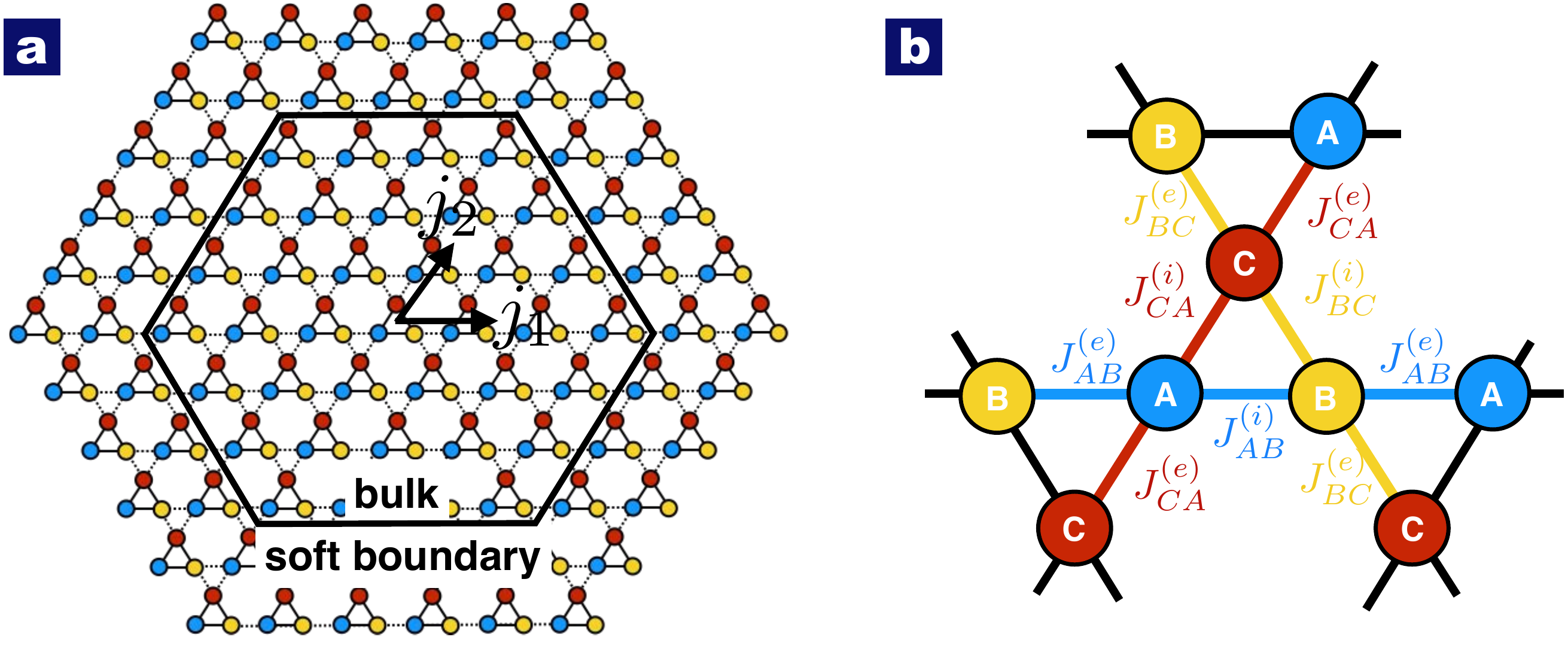}

\caption{\label{fig:latticemod}(a) Scheme of the Kagome tight binding lattice
on a hexagon. The solid triangles indicate the unit cells, whereas
the dashed lines illustrate the nearest neighbor coupling to different
unit cells. The three different colors illustrate the different sublattice
sites. The red hexagonal line separates the bulk area (inner part)
from the boundary area (outer part) (b) Zoom into the lattice, depicting
one unit cell and its six neighboring sites. Within one unit cell
mechanical excitations can hop from one site to another with the internal
hopping rates $J_{AB}^{(i)}$, $J_{BA}^{(i)}$ and $J_{CA}^{(i)}$,
whereas external hopping rates out of one unit cell are labeled with
$J_{AB}^{(e)}$, $J_{BA}^{(e)}$ and $J_{CA}^{(e)}$. Note that internal
and external hopping rates are equal if the $\mathcal{C}_{2}$-symmetry
is maintained.}
\end{figure}

For the transport calculations, we have modeled the hexagonal snowflake
crystal by a tight-binding Hamiltonian on a Kagome lattice, which
is the lattice that describes the links between neighboring triangles,
whose motion represents the relevant sound waves for the particular
triple of bands that we choose to consider. The Kagome lattice Hamiltonian
reads, 
\begin{equation}
\hat{H}=\sum_{\mathbf{j}}(\bar{E}+\delta E)\hat{a}_{\mathbf{j}}^{\dagger}\hat{a}_{\mathbf{j}}+\sum_{\langle\mathbf{j},\mathbf{j}'\rangle}J_{{\rm \mathbf{j,j'}}}\hat{a}_{\mathbf{j}}^{\dagger}\hat{a}_{\mathbf{j'}}.\label{eq:KagHam}
\end{equation}
Here, $\mathbf{j}$ is a multi-index, $\mathbf{j}=(j_{1},j_{2},s)$
where $j_{1/2}$ label the unit cell, and $s=A,B,C$ the sublattice,
see Figure \ref{fig:latticemod}a. As usual, $\langle\mathbf{j},\mathbf{j}'\rangle$
indicates the sum over the nearest neighbors. The hopping matrix $K_{{\rm \mathbf{j,j'}}}$
is symmetric and its matrix elements are chosen to reproduce the same
Dirac equation that would effectively describe our patterned snowflake
crystal. The energy $\bar{E}$ is the eigenenergy of the states $\boldsymbol{\psi}_{m_{s},\tau}$
for the rotationally symmetric crystal (see main text) while $\delta E$
cancel out a renormalization of the energy by the hopping terms. 

In the main text and in the Appendix \ref{sec:FEM}, we have shown
how the FEM simulations can be mapped onto the effective Dirac Hamiltonian
Equation (1) of the main text. Here, we show how the tight-binding
model Eq. (\ref{eq:KagHam}) can be mapped onto the same effective
Hamiltonian. We first consider the simple case where the invariance
under discrete translations and the ${\cal C}_{6}$ symmetry are not
broken corresponding to $m=0$ and $\mathbf{A}=0$. In this case,
all (nearest-neighbor) hopping rates must be equal $K_{{\rm \mathbf{j,j'}}}=K.$
One can easily calculate the equivalent first-quantized Hamiltonian
\[
h(\mathbf{k})=\begin{pmatrix}\bar{E}+\delta E & J(1+e^{-i\mathbf{k}\cdot\boldsymbol{a}_{1}}) & J(1+e^{i\mathbf{k}\cdot\boldsymbol{a}_{3}})\\
J(1+e^{i\mathbf{k}\cdot\boldsymbol{a}_{1}}) & \bar{E}+\delta E & J(1+e^{-i\mathbf{k}\cdot\boldsymbol{a}_{2}})\\
J(1+e^{-i\mathbf{k}\cdot\boldsymbol{a}_{3}}) & J(1+e^{i\mathbf{k}\cdot\boldsymbol{a}_{2}}) & \bar{E}+\delta E
\end{pmatrix},
\]
where $\boldsymbol{a}_{1}=(a,0),$ $\boldsymbol{a}_{2}=a(-1,\sqrt{3})/2,$
and $\boldsymbol{a}_{3}=a(-1,-\sqrt{3})/2$. By expanding around $\mathbf{k}=\mathbf{K}$
or $\mathbf{k}=\mathbf{K}'$ and projecting onto the states $|m_{1}=0,\tau\rangle=(1,1,1)/\sqrt{3}$
and $|m_{1}=-\tau,\tau\rangle=(1,e^{-i2\pi\tau/3},e^{i2\pi\tau/3})/\sqrt{3}$
($m_{1}$ is the quasimomentum about the center of the downwards triangles)
we find the Dirac Hamiltonian (1) of the main text with $m=0,$ $\mathbf{A}=0,$
and $v=\sqrt{3}aK/2$. The energy at the tip of the cones is $\bar{E}$
if $\delta E=-J$.

Next, we break the ${\cal C}_{6}$ symmetry but preserve the translational
symmetry such that the mass $m$ and the gauge $\mathbf{A}$ fields
are constants. In this case, there are six different hopping rates.
Three of them describe the hopping within the same (to a different)
unit cell $J_{\mathbf{j},\mathbf{j'}}=J_{s,s'}^{(i)}=J+\delta J_{s,s'}^{(i)}$
($J_{\mathbf{j},\mathbf{j'}}=J_{s,s'}^{(e)}=J+\delta J_{s,s'}^{(e)}$)
where $s\neq s'$ and $j_{x/y}=j_{x/y}'$ ($j_{x/y}\neq j_{x/y}'$).
The resulting first-quantized Hamiltonian reads
\[
h(\mathbf{k})=\begin{pmatrix}\bar{E}+\delta E & h_{AB}(\mathbf{k}\cdot\boldsymbol{a}_{1}) & h_{CA}^{*}(\mathbf{k}\cdot\boldsymbol{a}_{3})\\
h_{AB}^{*}(\mathbf{k}\cdot\boldsymbol{a}_{1}) & \bar{E}+\delta E & h_{BC}(\mathbf{k}\cdot\boldsymbol{a}_{2})\\
h_{CA}(\mathbf{k}\cdot\boldsymbol{a}_{3}) & h_{BC}^{*}(\mathbf{k}\cdot\boldsymbol{a}_{2}) & \bar{E}+\delta E
\end{pmatrix},
\]
where $h_{ss'}(\mathbf{x)}\equiv J_{ss'}^{(i)}+J_{ss'}^{(e)}e^{-i\mathbf{x}}.$
Since we are interested in the case where the ${\cal C}_{6}$ symmetry
is broken only weakly, we assume $\delta J_{s,s'}^{(i/e)}\ll J$.
Up to leading order in $\delta J_{s,s'}^{(i/e)}$ this correspond
to the the Dirac Hamiltonian (1) of the main text with
\begin{eqnarray}
\mathbf{A} & \approx & -\tau[(J_{AB}^{(e)}+J_{AB}^{(i)})\mathbf{e}_{1}+(J_{BC}^{(e)}+J_{BC}^{(i)})\mathbf{e}_{2}\nonumber \\
 &  & \quad\,\,+(J_{CA}^{(e)}+J_{CA}^{(i)})\mathbf{e}_{3}]/v,\label{eq:vectorpottight-binding}\\
m & \approx & (J_{AB}^{(e)}+J_{BC}^{(e)}+J_{CA}^{(e)}-J_{AB}^{(i)}-J_{BC}^{(i)}-J_{CA}^{(i)})/2,\,\,\,\,\,\,\,\,\,\,\,\,\label{eq:masstight-binding}\\
\delta E & \approx- & (J_{AB}^{(i)}+J_{BC}^{(i)}+J_{CA}^{(i)}+J_{AB}^{(e)}+J_{BC}^{(e)}+J_{CA}^{(e)})/6,\,\,\,\,\,\,\,\,\,\,\,\label{eq:onsitetight-binding}
\end{eqnarray}
 where the vectors $\mathbf{e}_{j}$ are defined in the main text. 

With the help of the relations (\ref{eq:vectorpottight-binding},\ref{eq:masstight-binding},\ref{eq:onsitetight-binding})
we can simulate the Dirac Hamiltonian (1) of the main text with the
desired constant pseudo-magnetic field (in the symmetric gauge $\mathbf{A}=\tau B(y,-x,0)/2$)
and mass profile. To simulate the disorder we add and additional random
energy shift $\Omega_{\mathbf{j}}$ equally distributed on the interval
$-\Delta\Omega/2<\Omega_{\mathbf{j}}<\Delta\Omega/2.$ A finite decay
rate of the phonons (necessary to reach a steady state), is described
within the standard input/output formalism \cite{GerryKnight}. In
the simulations with the drains, the decay rate is increase smoothly
in the regions of the drains (in order to avoid introducing any additional
backscattering).

\section{Possible Implementations\label{appG}}

Here, we provide a few more comments regarding the optomechanical
excitation and read out of helical waves. In the simplest approach
without a cavity, one could illuminate the structure from above by
tightly focussed laser beams, exerting radiation pressure directly.
A rough estimate of the force, for a laser power of $1\,\text{mW}$,
indicates that (out-of-plane) vibrational amplitudes of the order
of $10$fm might be achieved. In this estimate, we have adopted the
simplest possible approach, treating the triangle as an oscillator
with a frequency of order $2\pi\cdot14{\rm GHz}$ and a decay rate
($2{\rm GHz}$) set by the scale of the bandwidth of the Kagome bands.
A more detailed analysis would be needed to extract the excitation
efficiency for the particular vibrational modes of interest, which
formed the basis of our discussion in the main text.

\begin{figure}
\centering{}\includegraphics[width=1\columnwidth]{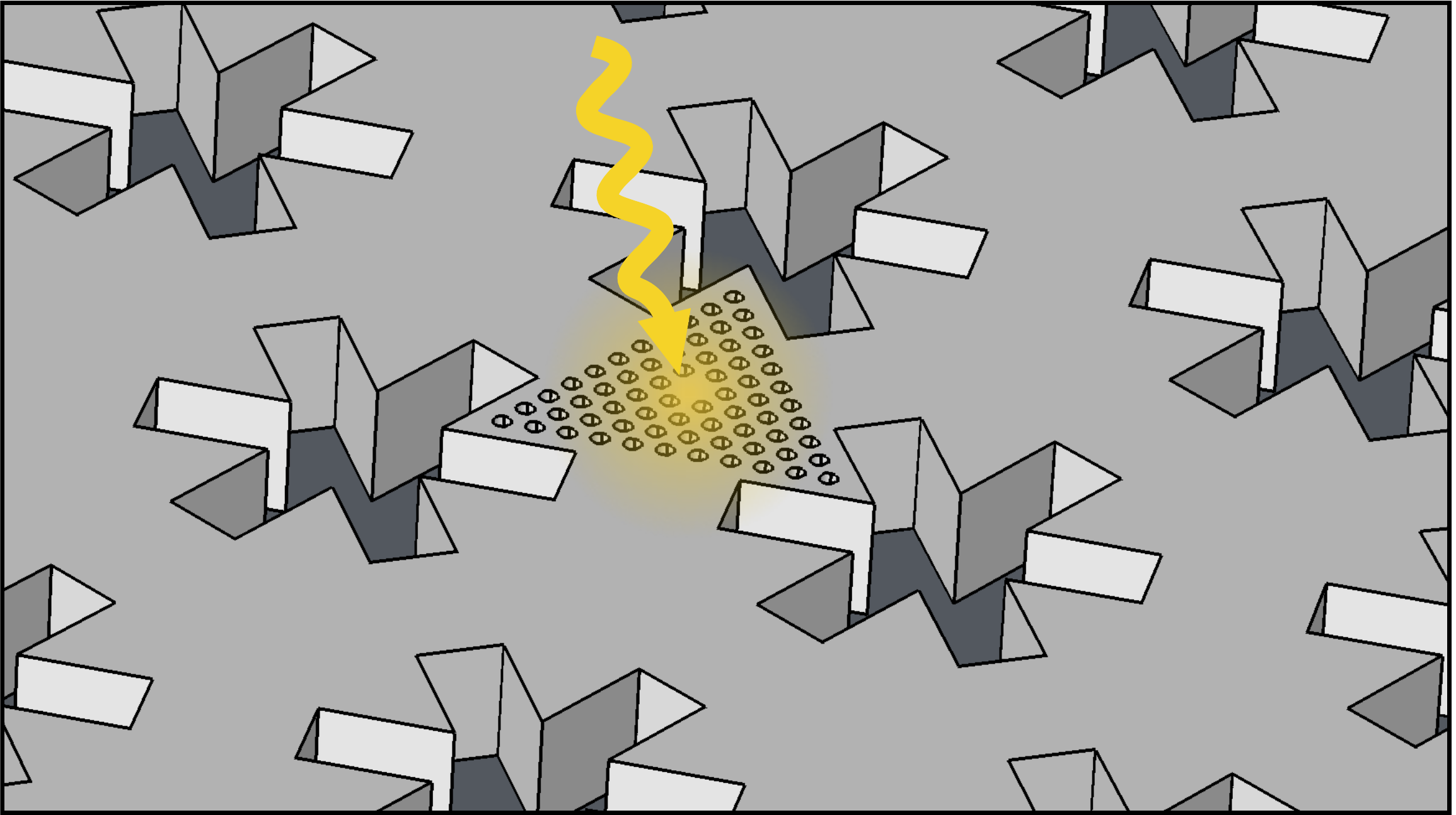}\caption{\label{fig:CavOnTri}Schematic picture of a scaled up snowflake array
with one triangle hosting a defect-mode nano cavity. Impinging light
is enhanced by the cavity's finesse thereby launching mechanical excitations
via the optomechanical interaction.}
\end{figure}

However, a much more efficient approach is available, involving optical
cavities. A structure scaled up by a factor of $X=10$ (resulting
in frequencies lower by $X$) can host defect-mode nano cavities embedded
in the triangles itself, cf. Figure \ref{fig:CavOnTri}. For any such
optical cavity, a circulating light with a modulated intensity will
give rise (via radiation pressure and photoelastic forces) to periodic
cycles of expansion and contraction of the structure. This type of
motion (when the right frequency is selected) clearly overlaps with
the vibrational modes that are relevant for our proposal, cf. Fig.
1d\textendash f of the main text. We note that the light intensity
is enhanced by the cavity's finesse (usually at least $\gtrsim100$),
thereby increasing the amplitude of the vibrations. As the thermal
motion decreases with the factor $1/\sqrt{X}$, it is easily possible
to overcome the thermal motion at room temperature. In addition, during
the measurements one can average out the noise and provide a clear
signal of the excited sound waves traveling through the structure,
regardless of thermal fluctuations. 

\end{document}